\documentclass[aip,reprint]{revtex4-1}
\usepackage{graphicx}
\usepackage{amsmath}
\usepackage{mathptmx}
\usepackage{eucal}

\draft 

\begin{document}

\title{Fundamental Bounds of Wavefront Shaping of Spatially Entangled Photons} 

\author{Ronen Shekel}
\affiliation{Racah Institute of Physics,The Hebrew University of Jerusalem, Jerusalem, 91904, Israel}
\author{Sébastien M. Popoff}
\affiliation{Institut Langevin, ESPCI Paris, PSL University, CNRS, France}
\author{Yaron Bromberg}
\email[]{Yaron.Bromberg@mail.huji.ac.il}
\affiliation{Racah Institute of Physics,The Hebrew University of Jerusalem, Jerusalem, 91904, Israel}

\date{\today}

\begin{abstract}
Wavefront shaping enables control of classical light through scattering media. Extending these techniques to spatially entangled photons promises new quantum applications, but their fundamental limits, especially when both photons scatter, remain unclear. Here, we theoretically and numerically investigate the enhancement of two-photon correlations through thick scattering media. We analyze configurations where a spatial light modulator shapes one or both photons, either before or after the medium, and show that the optimal enhancement differs fundamentally from classical expectations. For a system with $N$ modes, we show that shaping one photon yields the classical enhancement $\eta \approx (\pi/4)N$, while shaping both photons before the medium reduces it to $\eta \approx (\pi/4)^2N$. However, in some symmetric detection schemes, when both photons are measured at the same mode, perfect correlations are restored with $\eta \approx N$, resembling digital optical phase conjugation. Conversely, shaping both photons after the medium leads to a complex, NP-hard-like optimization problem, yet achieves superior enhancements, up to $\eta \approx 4.6N$. These results reveal unique quantum effects in complex media and identify strategies for quantum imaging and communication through scattering environments.
\end{abstract}

\pacs{}

\maketitle 

\section{Introduction}
When classical light propagates through a complex medium such as white paint or biological tissue, it is scattered, producing a random speckle pattern. In their seminal work ~\cite{vellekoop2007focusing}, Vellekoop and Mosk demonstrated that by shaping the incoming wavefront using a spatial light modulator (SLM), the strong scattering could be overcome, initiating the field of wavefront shaping ~\cite{mosk2012controlling}. Since then, the scope of wavefront shaping has grown considerably, with advances in optimization algorithms, feedback mechanisms, and a variety of shaping modalities ~\cite{popoff2010measuring, vellekoop2015feedback, cao2022shaping, gigan2022roadmap}. 

While wavefront shaping in the classical domain continues to be a powerful technique for applications ranging from deep-tissue imaging to optical communications, the extension of these concepts to the quantum domain ~\cite{lib2022quantum} opens new possibilities in areas such as quantum imaging and communications ~\cite{goyal2016effect, ndagano2017characterizing, liu2019single, cao2020long, amitonova2020quantum, pugh2020adaptive}. In particular, spatially entangled photons produced by spontaneous parametric down-conversion (SPDC) exhibit strong correlations that are also scrambled when propagating through a complex medium, resulting with a two-photon speckle ~\cite{beenakker2009two, peeters2010observation}. Over the past decade, experiments have shown that standard wavefront shaping tools can be applied directly to these two-photon correlations. 

In quantum communication schemes, one photon of an entangled pair is typically retained at the transmitter while the other is sent through a complex quantum channel. Several works~\cite{carpenter2013mode,defienne2014nonclassical,huisman2014controlling,valencia2020unscrambling, shekel2023pianoq, devaux2023restoring} have studied this scenario, where only a single photon experiences scattering. Since classical and quantum light share the same spatial modes, the scattering of classical light, a weak coherent beam, and heralded single photons are all described by the same transmission matrix, and a classical beacon could be used to find optimal phases that restore the correlations ~\cite{carpenter2013mode,defienne2014nonclassical,huisman2014controlling, devaux2023restoring}. 

In quantum imaging and related applications, both photons may propagate through the complex medium~\cite{defienne2016two,pinkse2016programmable,defienne2018adaptive, Peng2018manipulation, lib2020real,lib2020pump, soro2021quantum, cameron2024adaptive, shekel2024shaping,courme2025non}. In this case, the phases that localize the correlations after the scattering differ from those that would focus a classical beacon beam ~\cite{courme2025non}, and therefore a classical beacon cannot be used for feedback. One solution is to use the quantum signal itself for feedback \cite{Peng2018manipulation,shekel2023pianoq,cameron2024adaptive,courme2025non}. Alternatively, several classical feedback strategies have been proposed: (a) measuring the one-photon transmission matrix and inferring from it the two-photon transmission matrix ~\cite{defienne2016two,pinkse2016programmable}; (b) using the pump beam for feedback in the case of thin diffusers ~\cite{lib2020real, lib2020pump, shekel2021shaping}; or (c) employing an advanced wave beacon beam ~\cite{shekel2024shaping}. The fact that a classical beacon scatters differently from the two-photon state implies that the performance bounds of wavefront shaping, derived for classical light~\cite{vellekoop2008phase}, are not directly applicable to this quantum scenario.

In this work, we examine the fundamental limits of restoring the correlations between entangled photons that both propagate through a thick scattering medium. We show, both analytically and numerically, that the optimal enhancement in the quantum case differs from the classical enhancement and depends on several factors, including the relative locations of the two detectors and the plane at which the SLM is positioned. In particular, we show that in certain symmetric scenarios \textit{perfect} compensation of scattering is achievable, an effect we interpret as a form of digital optical phase conjugation. We further show that when the SLM is placed after the scattering medium, the optimization problem resembles NP-hard problems. Nevertheless, numerical solutions exhibit superior performance compared to classical shaping. Notably, this includes instances where the optimized coincidence rate at the target modes exceeds the total coincidence rate summed over all modes before shaping. These results are determined by the optimal shaping phases, regardless of the method used to obtain them.

\section{Results}
We consider spatially entangled photons generated via spontaneous parametric down-conversion (SPDC) ~\cite{walborn2010spatial}. In SPDC, a strong pump beam impinges on a nonlinear crystal, and with some probability, one pump photon is annihilated and two lower-energy photons are created. Depending on the phase-matching conditions of the nonlinear crystal, the generated photon pairs can be entangled in their polarization, spectral, and spatial degrees of freedom. Since this work focuses on spatial wavefront shaping, we simplify the analysis by considering the case where the photons share the same polarization (type-I SPDC) and the same frequency (degenerate SPDC). Under these conditions, we examine spatially entangled photons, whose state can be approximated by $\left|\Psi_{\text{in}}\right\rangle =\frac{1}{\sqrt{2N}}\sum_{\textbf{x}}\hat{a}_{\textbf{x}}^{\dagger}\hat{a}_{\textbf{x}}^{\dagger}\left|\text{vac}\right\rangle$, where $\hat{a}_{\textbf{x}}^{\dagger}$ is the creation operator in the transverse position $\textbf{x}$, $\left|\text{vac}\right\rangle$ denotes the vacuum state and $N$ is the number of modes in the discretized space. Experimentally, this state is obtained by weakly focusing the pump beam on the nonlinear crystal, operating in the so-called \textit{thin crystal regime} ~\cite{howell2004realization, walborn2010spatial}.

We are interested in a scenario where both photons propagate through a thick scattering sample. We model the scattering medium by a transmission matrix $T$ ~\cite{popoff2010measuring}, and consider two models for $T$: i) A random unitary matrix, which describes, for example, a lossless multimode fiber with strong mode mixing. ii) A matrix whose elements are independently and identically distributed random variables drawn from a circular complex Gaussian distribution (Gaussian IID)~\cite{goodman2007speckle}, representing a subspace of a strongly scattering medium ~\cite{goetschy2013filtering}. 

\begin{figure}[ht!]
    \centering
    \includegraphics[width=\linewidth]{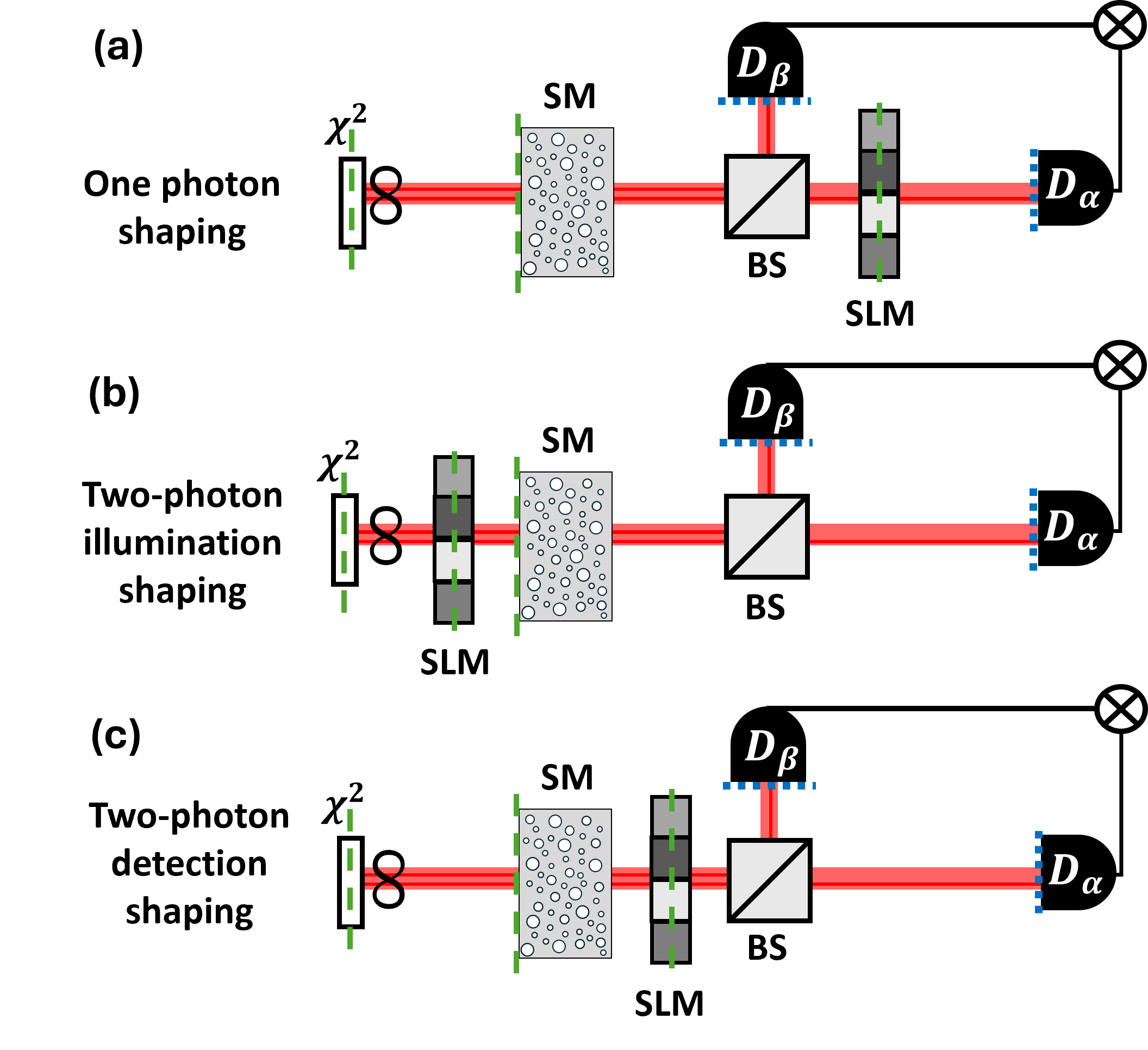} 
    \caption{\label{fig:SPDC_configs}The three configurations for wavefront shaping of spatially entangled photons analyzed in this work. In all configurations entangled photons are generated in a $\chi^{(2)}$ SPDC process (type-I), and then propagate through a scattering medium (SM). Their correlations are detected using coincidence events between two single photon detectors ($D_\alpha, D_\beta$).  (a) The one-photon shaping configuration, where only one of the photons is shaped. (b) The two-photon illumination shaping configuration, where the SLM shapes both photons before propagating through the scattering medium. (c) The two-photon detection shaping configuration, where the SLM is positioned after the scattering sample, and shapes both photons. The green dashed lines mark conjugate planes, and the dotted blue lines mark the far-field of these planes.}
\end{figure}

As depicted in Fig. \ref{fig:SPDC_configs}, we analyze three different configurations for positioning an SLM to compensate for this scattering. In the first configuration (Fig. \ref{fig:SPDC_configs}a), the SLM is placed after the scattering medium and shapes only one of the photons. We refer to this configuration as one-photon shaping (1P-S). In the second configuration (Fig. \ref{fig:SPDC_configs}b), the SLM is placed before the scattering medium and shapes both photons. We call this configuration two-photon illumination shaping (2P-IS). In the third configuration (Fig. \ref{fig:SPDC_configs}c), the SLM is placed after the scattering medium and shapes both photons. We refer to it as two-photon detection shaping (2P-DS). In all three configurations, we assume the SLM is placed in the image plane of the input facet of the scattering medium and of the nonlinear crystal, such that the two photons illuminate the same area on the sample, and the detectors are placed at its far-field.  

The goal of the wavefront shaping schemes we consider is to maximize the probability to find the photons in two spatial modes $\alpha$ and $\beta$, corresponding to a coincidence event. For the three configurations we consider, this probability is given by (see Supplementary Information Section S1): 

\begin{equation} \label{eq:coin}
    \begin{aligned}
        P^{(1P-S)}_{\alpha\beta} &= \frac{1}{2N} \left| \left(\mathcal{F}TT^{\mathsf{T}}S\mathcal{F} \right)_{\beta\alpha} \right|^2
        \\
        P^{(2P-IS)}_{\alpha\beta} &= \frac{1}{2N} \left| \left(\mathcal{F}TSST^{\mathsf{T}}\mathcal{F} \right)_{\beta\alpha} \right|^2 
        \\
        P^{(2P-DS)}_{\alpha\beta} &= \frac{1}{2N} \left|\left(\mathcal{F}STT^{\mathsf{T}}S\mathcal{F} \right)_{\beta\alpha} \right|^2,
    \end{aligned}
\end{equation}

where $S$ is a diagonal matrix representing the operation performed by the SLM in the pixel basis, and $\mathcal{F}$ is the discrete Fourier transform operator, representing the propagation to the detectors plane in the far-field.   

\begin{figure}[ht!]
    \centering
    \includegraphics[width=\linewidth]{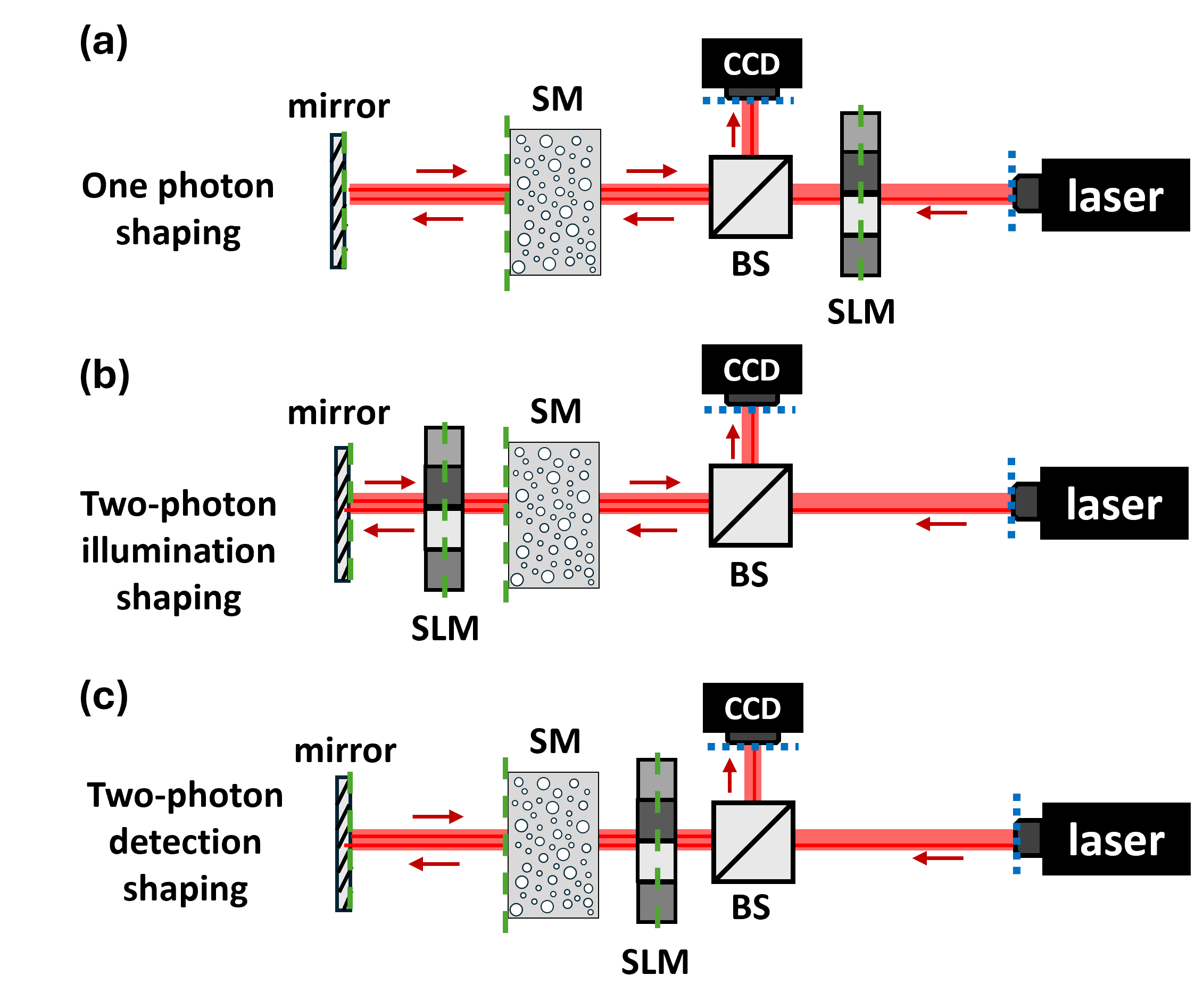} 
    \caption{\label{fig:Klyshko_configs}Depiction of the three analogous classical setups according to the advanced wave picture. (a) The analogous 1P-S configuration, which is a classical wavefront shaping scenario in reflection geometry. (b) The analogous 2P-IS configuration, where the light is scattered both before and after being shaped by the SLM. (c) The analogous 2P-DS configuration, involving light impinging on the same SLM twice: before and after the scattering media. The green dashed lines mark conjugate planes, and the dotted blue lines mark their far-field plane.}
\end{figure}

To gain further physical intuition into these three configurations, it is useful to interpret them using Klyshko's advanced wave picture ~\cite{klyshko1988simple, belinskii1994two, shekel2024shaping, zheng2024AWP}. In this picture, the probability $P$ for a coincidence event between two single-photon detectors in a two-photon setup is translated to a classical intensity measurement, by substituting one of the detectors with a classical light source and the nonlinear crystal with a mirror. The light emitted from this source is directed back through the optical setup toward the mirror, and is then reflected to the second detector. According to the advanced wave picture, the intensity measured at the second detector is proportional to the average coincidence rate observed in the original two-photon experiment. In Fig. \ref{fig:Klyshko_configs}, we show the advanced wave interpretation of the three configurations described in Fig.~\ref{fig:SPDC_configs}. While the calculations yield the same expressions in both pictures (see Supplementary Information Section S1), the advanced wave picture provides several physical insights that help interpret the three configurations. 

\subsection{One photon shaping (1P-S)}
We begin by analyzing the 1P-S configuration, where only one of the SPDC photons is shaped. As shown by the advanced wave interpretation (Fig. \ref{fig:Klyshko_configs}a), this case is equivalent to a standard classical wavefront-shaping experiment in a reflection geometry, where light traverses the medium twice. The problem thus reduces to maximizing the intensity of a classical coherent field emitted from mode $\alpha$ and detected at mode $\beta$. 

The figure of merit we use to define the optimization quality is the enhancement factor $\eta$. This is defined as the ratio between the maximal coincidence probability $P_{\alpha\beta}$, achieved by optimizing the SLM phases, and the spatially-averaged coincidence probability before optimization. As we show explicitly in the Supplementary Information Section S2, for 1P-S and a phase-only SLM, the enhancement is given by:

\begin{equation}
    \eta_{1P-S} = 1 + \left(N-1\right) \frac{\pi}{4},
\end{equation}

where $N$ denotes the number of modes in the system. This reproduces the well‑known classical result of linear scaling in $N$ with a pre-factor of $\frac{\pi}{4}$ for phase‑only control ~\cite{vellekoop2008phase}.

While the enhancement is the same for the unitary and Gaussian IID cases, the peak-to-background ratio is different. In the $N\gg1$ limit, the peak in both cases is $\eta/N\approx\pi/4$. However, in the unitary case, the background is $1-\eta/N$, since the peak gets its power from the background, while for Gaussian IID, the total background power is higher. 


\subsection{Two-photon illumination shaping (2P-IS)}
We now consider the 2P-IS configuration (Fig. \ref{fig:SPDC_configs}b), where the SLM shapes both photons before the scattering medium. The SLM is typically placed in an image plane of the crystal to ensure the phase mask $S$ acts identically on both photons, as described in Eq. \ref{eq:coin} by $SS$. In the advanced wave counterpart (Fig. \ref{fig:Klyshko_configs}b), this corresponds to a configuration in which the light is scattered both before and after passing through the SLM. Here, the fact that the SLM is in the image plane of the crystal implies that the light hits the SLM twice with the same spatial distribution, effectively equivalent to a single pass where each pixel induces a double phase.

By performing an analytical calculation similar to that of the 1P-S configuration (see Supplementary Information Section S2), we obtain:

\begin{equation}
    \begin{aligned}
        \eta_{2P-IS} = 1+\left(N-1\right)\left(\frac{\pi}{4}\right)^2.
    \end{aligned}
\end{equation}
    
For $N \gg 1$, the enhancement is lower by a factor of $\pi/4$ compared to $\eta_{1P-S}$. This can be intuitively understood in the advanced wave picture: in this configuration, the SLM is illuminated with a speckle pattern with varying amplitude, so that some of its $N$ pixels contribute less effectively to the optimization, leading to a reduced number of effective degrees of control. 

\subsubsection*{Enhanced correlations by digital optical phase conjugation}
Although the 2P-IS configuration typically results in a lower enhancement, a remarkable effect occurs when optimizing the probability $P_{\alpha\alpha}$ that both photons are detected in the \textit{same} spatial mode, namely, when the transverse positions of the two detectors are identical. In this symmetric case, the two-photon correlation enhancement is given by 

\begin{equation}
    \eta_{2P-IS}^{(OPC)}=N + b,
\end{equation}

where $b=0$ for the unitary case and $b=1$ for the Gaussian IID case. Remarkably, the enhancement scales linearly with the number of modes with a pre-factor of exactly 1, despite the use of phase-only modulation. Before optimization, upon detection of a photon in some mode $\alpha$, the probability to detect its twin photon in some specific mode is, on average, $1/N$. After optimization, the probability to detect the twin photon in mode $\alpha$ becomes unity. Notably, in the unitary case there is no residual background ($1-\eta/N=0$), manifesting the \textit{perfect} enhancement achieved. Further details are given in the Supplementary Information Section S2. 

It is instructive to interpret the perfect correlations obtained when the two photons are detected in the same mode, as digital optical phase conjugation in the advanced wave picture ~\cite{cui2010implementation}. Perfect correlations, which correspond in the advanced wave picture to focusing all the light back towards the source, are obtained when the phases applied by the SLM are the conjugate of the field incident on it after passing through the scattering sample (Fig.~\ref{fig:Klyshko_configs}b). In this case, the field after the SLM becomes the phase-conjugate of the incident field. By time-reversal symmetry, this results in all the light being focused back towards the sources ~\cite{boyd2008nonlinear}. 

In contrast to typical digital optical phase conjugation configurations, where perfect time reversal requires precise control over both the amplitude and phase of the scattered field, here perfect time reversal is achieved using phase conjugation alone. This is possible because, in the advanced wave picture, the SLM is placed after the scattering medium. As a result, the amplitude of the field illuminating the SLM is already correct, and only the phases need to be conjugated. We emphasize that the SLM appears after the scattering medium only in the advanced wave picture. In practice, the SLM shapes the two photons before they illuminate the sample. We refer to this special configuration as 2P-IS(OPC). 

\subsection{Two-photon detection shaping (2P-DS)}
Next, we consider the 2P-DS configuration, where the SLM shapes the two photons \textit{after} the scattering medium (Fig. \ref{fig:SPDC_configs}c). In the advanced wave picture (Fig. \ref{fig:Klyshko_configs}c), the light now encounters the same SLM twice, with random scattering occurring in between the two encounters. 

The advanced wave picture highlights that the 2P-DS poses a significant optimization challenge: attempting to optimize the SLM phases using standard iterative algorithms ~\cite{vellekoop2015feedback} fails because the SLM pixels are no longer independent. The phase applied to one pixel affects not only the light that is reflected from that pixel but also the light that will impinge on other pixels when the beam illuminates the SLM for the second time. This is similar to the result reported in ~\cite{courme2025non}, where it was found that the optimization landscape for entangled photon correlations is non-convex. In the Supplementary Information Section S3 we show that maximizing the two-photon correlations in the 2P-DS case can be mapped to a version of a maximum quadratic program (MAXQP) or of finding the ground state of an XY Ising model, which are known to be NP-hard problems ~\cite{arora2005non, pardalos1991quadratic, lucas2014ising, pierangeli2019large, pierangeli2021scalable, courme2025non}. 

Despite this computational complexity, we use PyTorch's autograd engine for gradient-based optimization of phase patterns that localize the correlations between two target modes, in a system with $N=512$ modes. Full details are provided in the Supplementary Information Section S4 and in the supplementary code ~\cite{qwfs2024}. Unlike the 1P-S and 2P-IS configurations, in this scenario our simulations reveal distinct behavior for unitary and Gaussian IID transmission matrices: For the unitary case, we find a value of $\eta_{2p-DS}^{(U)}\approx0.89 \cdot N$, and for the Gaussian IID case, our optimization yields a remarkable enhancement of $\eta_{2p-DS}^{(G)}\approx 1.91\cdot N$, such that the coincidence rate between the optimized modes exceeds the total coincidence rate before optimization. This is possible, since the shaping process could redirect photons that would otherwise be scattered into unmeasured modes, such as reflected channels in the case of a scattering medium, into the measured ones. In both cases, the enhancement pre-factor slightly increases with the system size $N$, as we discuss in the Supplementary Information Section S4. 

Self-consistent optimization is not a unique feature of the 2P-DS configuration and may also arise in the 2P-IS configuration, for instance, when the SLM is not imaged onto the nonlinear crystal, as we discuss in the Supplementary Information Section S5.

\subsubsection*{Enhanced correlations by digital optical phase conjugation}
Similarly to the 2P-IS configuration, also in detection shaping we observe that maximizing the probability that both photons arrive at the same detector increases the enhancement factor. For the unitary case, we once again numerically achieve a perfect enhancement of $\eta_{2P-DS}^{(OPC)}=N$. We interpret this as a more intricate version of digital optical phase conjugation in the advanced wave picture (Fig. \ref{fig:Klyshko_configs}c). In this picture, for the reflected field to be the time-reversed version of the incident field, the reflection process at the mirror must correspond to phase conjugation. First, we note that if the phases arrive at the mirror plane with a flat phase, it will effectively act as a phase-conjugating mirror, since the phase conjugation of a flat phase is a flat phase. Similarly, if the field in the mirror plane is real valued, namely the phases are only $0$ or $\pi$, the mirror effectively acts as a phase-conjugating mirror. Thus, the optimal phases that the SLM finds in this configuration are such that the field at the mirror plane is real (see Fig. S5). Since there are two possible phases for each mode, there are $2^{N}$ such solutions, and it is thus relatively simple for the optimization process to find such a solution and achieve perfect correlations. We refer to this special configuration as 2P-DS(OPC). 

For Gaussian IID transmission matrices, the enhancement is not bounded from above by $N$, and we find numerically an enhancement of $\eta_{2P-DS}^{(OPC)}\approx 4.6 \cdot N$. While we do not have a physical explanation for the enhancement value in the Gaussian IID case, we note that it is bounded by $\sigma_{1}^2$, where $\sigma_{1}$ is the maximal singular value of $TT^T$. Physically, $\sigma_1^2$ corresponds to the maximal intensity transmission through the medium ~\cite{goetschy2013filtering}.

\begin{figure*}[htpb!]
    \centering
    \includegraphics[width=\linewidth]{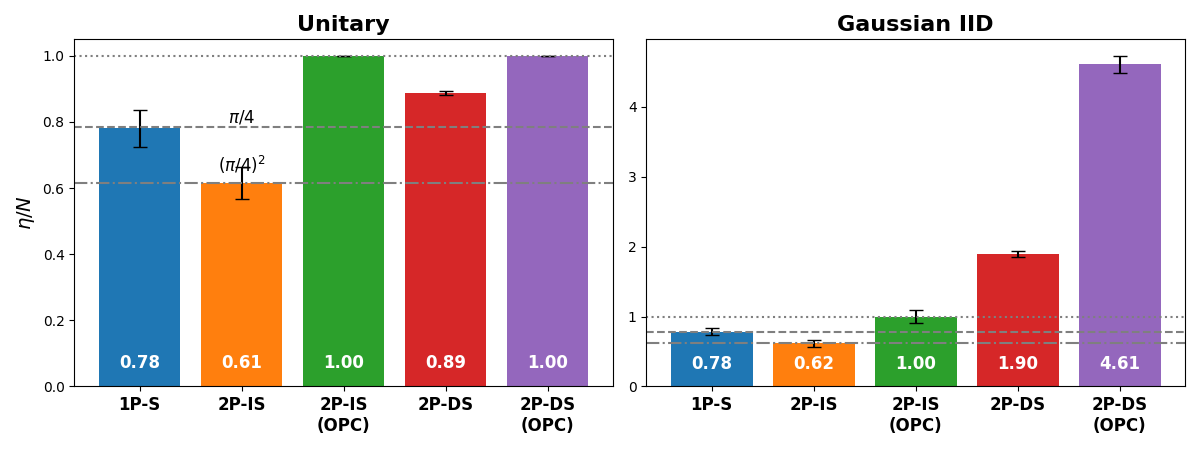} 
    \caption{\label{fig:summary}Summary of the enhancement pre-factor for the five different analyzed configurations. The enhancements were calculated numerically by averaging over 200 disorder realizations, for a system with 512 modes. The errorbars signify the standard deviation of the 200 realizations. The 1P-S, 2P-IS, and 2P-IS(OPC) configurations present the same results for the Unitary and Gaussian IID cases, and agree with the analytical calculations of $\pi/4$, $\left(\pi/4\right)^2$, and $1$, respectively. The 2P-DS and 2P-DS(OPC) configurations are significantly different for the unitary and Gaussian IID cases. In the latter, the coincidence rate between the optimized modes exceeds the total coincidence rate prior to optimization.}
\end{figure*}

We summarize the enhancement pre-factors of the different configurations in Fig. \ref{fig:summary}. Notably, it highlights that the 2P-DS configuration puts forward a unique problem, where on one hand the optimization is hard and self-consistent, but on the other hand holds the potential for greatly improved performance, particularly in the symmetric configuration. To complement these results, in the Supplementary Information Sections S6 we also analyze the case of a thin diffuser. 

\subsection{Incomplete control} \label{sec:incomplete_control}
\begin{figure*}[htpb!]
    \centering
    \includegraphics[width=\linewidth]{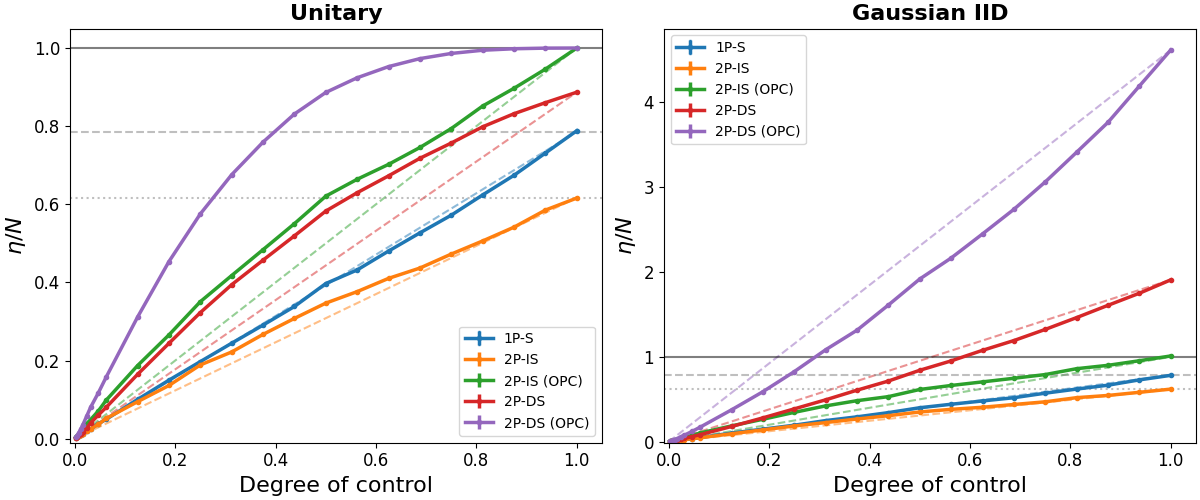} 
    \caption{\label{fig:incomplete_control}Scaling of the enhancement pre-factor as a function of the degree of control, defined as the ratio $\text{DOC} = M/N$ between the number of SLM pixels $M$ and the number of modes in the system $N$, for the analyzed configurations with unitary and Gaussian IID matrices. Dashed straight lines connect the endpoints for each curve to guide the eye regarding linear or nonlinear scaling. The results are shown for $N=512$.}
\end{figure*}

In most practical applications, the number of modes controllable via the SLM is limited and is often less than the total number of modes supported by the scattering medium. We denote this as \textit{incomplete control} and simulate how it influences the achievable enhancement $\eta$ in the different configurations. We define the degree of control (DOC) as the ratio between the number of controlled modes $M$ (e.g., independent SLM pixels) and the total number of modes considered in the system $N$. Figure \ref{fig:incomplete_control} illustrates the scaling of the enhancement pre-factor $\eta/N$ as a function of the degree of control for unitary and Gaussian IID transmission matrices.

First, we analyze the unitary case. For the 1P-S configuration, the enhancement exhibits a linear dependence on the DOC, consistent with classical wavefront shaping results where enhancement scales linearly with the number of controlled modes (see Supplementary Information Section S6). 

The 2P-IS configuration displays nonlinear scaling. As shown analytically in the Supplementary Information Section S6, the initial slope at low DOC is $\pi/4$. Since the enhancement pre-factor reaches $\left(\pi/4\right)^2$ at full control (DOC=1), the curve shows a decreasing slope, indicating a diminishing gain in enhancement as the degree of control increases. Intuitively, at a low degree of control, the SLM has large effective macro-pixels that average out the speckle illumination, effectively restoring a uniform beam condition akin to the 1P-S configuration.

The symmetric 2P-IS(OPC) configuration also shows nonlinear scaling with a decreasing slope. As derived in the Supplementary Information, the initial slope at a low degree of control is $\pi/2\approx 1.57$, while the enhancement pre-factor reaches $1$ at full control. This initial slope is notably twice the $\pi/4$ slope observed at low degrees of control for the non-symmetric 2P-IS case. This factor-of-two difference in the initial enhancement rate is reminiscent of coherent backscattering (CBS) ~\cite{akkermans2007mesoscopic, carminati2021principles}. Standard CBS describes the enhanced average intensity backreflected into the incident mode from a disordered medium, arising without any wavefront shaping, due to constructive interference between pairs of reciprocal paths inside the sample. Here, this constructive interference results in a two-fold improvement of the enhancement factor at low degrees of control, since in the advanced wave picture, the 2P-IS(OPC) configuration corresponds to a CBS setting ~\cite{lib2022thermal}.

Considering the unitary case, the 2P-DS configurations also exhibit nonlinear scaling with a decreasing slope. For the symmetric OPC case, this particularly efficient approach to high enhancement can be intuitively understood by recalling that there are $2^{N}$ valid options for phases in the mirror plane that will all result in a perfect enhancement. Thus, even when the control is incomplete, the existence of this vast solution space allows a fraction of these solutions to remain approximately reachable, leading to the observed saturation behavior at near-maximal enhancement even for DOC significantly less than 1 (see Fig. S8).

Next, we consider the Gaussian IID case. The 1P-S, 2P-IS, and 2P-IS(OPC) configurations exhibit scaling behaviors very similar to their unitary counterparts: linear for 1P-S, and nonlinear with a decreasing slope for both 2P-IS cases. However, the 2P-DS configurations behave distinctly in the Gaussian IID case, showing nonlinear scaling with an increasing slope. 

Even though we do not have an analytic model for the 2P-DS configuration, we empirically perform a linear fit to the slope of these configurations in the regime of low degree of control (see Fig. S7). We observe that on both unitary and Gaussian IID case, the initial slope of the symmetric 2P-DS(OPC) configuration is approximately twice that of the non-symmetric counterpart. This consistent factor-of-two difference reinforces the analogy to an extended CBS-like effect influencing the initial optimization efficiency.

\section{Discussion}
In this work, we have shown that, unlike in classical wavefront shaping, the specific placement and configuration of the SLM and detectors play a crucial role in determining the achievable enhancement when shaping entangled photons that both propagate through a thick scattering medium. Beyond their intrinsic interest, configurations where both photons traverse the medium also serve as a platform for applications that use complex media to implement tailored quantum circuits ~\cite{pinkse2016programmable, leedumrongwatthanakun2020programmable, goel2024inversedesign, valencia2024multiplexed}, and to study fundamental aspects of quantum light in complex media ~\cite{lahini2010quantum, black2019quantum, lib2022thermal, safadi2023coherent, bajar2025partial}. 

We found that two-photon wavefront shaping gives rise to a rich variety of behaviors with no direct counterpart in the classical regime. Depending on the configuration, two-photon correlations can either enhance or reduce performance compared to classical shaping and can exhibit fundamentally new features. For example, in the 2P-IS configuration, a natural symmetry of the two-photon measurement enables perfect compensation of scattering when focusing both photons to the same spatial mode, despite relying on phase-only control. 


In the 2P-DS configuration, where photons impinge twice on the SLM, leading to correlated pixel control, we have found a significant boost in enhancement. This improvement, however, comes at the price of a fundamentally more challenging, self-consistent optimization problem. Similar self-consistent optimization problems arise in advanced classical systems, such as coherent confocal microscopy through scattering media ~\cite{monin2025rapid} and tunable metasurface reflect-arrays inside chaotic microwave cavities ~\cite{del2018leveraging}. Interestingly, while presenting a challenge for optimization, the interdependence of pixels has been leveraged in optical information processing and neural networks, enabling nonlinear operations using linear optics ~\cite{xia2024nonlinear, yildirim2024nonlinear}. 

The emergence of complex feedback mechanisms in two-photon coincidence measurements highlights new opportunities to exploit quantum correlations, made even more promising by the robustness of the observed performance under incomplete control. 

We believe that these findings provide important insights into the foundations of quantum wavefront shaping and offer practical considerations for shaping entangled photons through complex media, with potential applications ranging from quantum imaging to secure communications.

\begin{acknowledgments}
The authors would like to thank Ohad Lib, Anat Levin and Ori Katz for many fruitful discussions. 
\end{acknowledgments}

\section*{Funding}
This research was supported by the Zuckerman STEM Leadership Program, the Israel Science Foundation (grant No. 2497/21). R.S. acknowledges the support of the Israeli Council for Higher Education and of the HUJI center for nanoscience and nanotechnology. S.M.P acknowledges the French \textit{Agence Nationale pour la Recherche} grant No. ANR-23-CE42-0010-01 and the Labex WIFI grant No. ANR-10-LABX-24, ANR-10-IDEX-0001-02 PSL*.

\section*{Data Availability Statement}
Code and data underlying the results presented in this paper are available in Ref. ~\cite{qwfs2024}. 

\section*{Disclosures}
The authors declare no conflict of interest.

\section*{Supplementary information}
Analytical derivations, exploration of other configurations, and further details regarding the numerical simulations may be found in the attached document. 

\bibliography{main}

\begin{thebibliography}{62}%
\makeatletter
\providecommand \@ifxundefined [1]{%
 \@ifx{#1\undefined}
}%
\providecommand \@ifnum [1]{%
 \ifnum #1\expandafter \@firstoftwo
 \else \expandafter \@secondoftwo
 \fi
}%
\providecommand \@ifx [1]{%
 \ifx #1\expandafter \@firstoftwo
 \else \expandafter \@secondoftwo
 \fi
}%
\providecommand \natexlab [1]{#1}%
\providecommand \enquote  [1]{``#1''}%
\providecommand \bibnamefont  [1]{#1}%
\providecommand \bibfnamefont [1]{#1}%
\providecommand \citenamefont [1]{#1}%
\providecommand \href@noop [0]{\@secondoftwo}%
\providecommand \href [0]{\begingroup \@sanitize@url \@href}%
\providecommand \@href[1]{\@@startlink{#1}\@@href}%
\providecommand \@@href[1]{\endgroup#1\@@endlink}%
\providecommand \@sanitize@url [0]{\catcode `\\12\catcode `\$12\catcode `\&12\catcode `\#12\catcode `\^12\catcode `\_12\catcode `\%12\relax}%
\providecommand \@@startlink[1]{}%
\providecommand \@@endlink[0]{}%
\providecommand \url  [0]{\begingroup\@sanitize@url \@url }%
\providecommand \@url [1]{\endgroup\@href {#1}{\urlprefix }}%
\providecommand \urlprefix  [0]{URL }%
\providecommand \Eprint [0]{\href }%
\providecommand \doibase [0]{http://dx.doi.org/}%
\providecommand \selectlanguage [0]{\@gobble}%
\providecommand \bibinfo  [0]{\@secondoftwo}%
\providecommand \bibfield  [0]{\@secondoftwo}%
\providecommand \translation [1]{[#1]}%
\providecommand \BibitemOpen [0]{}%
\providecommand \bibitemStop [0]{}%
\providecommand \bibitemNoStop [0]{.\EOS\space}%
\providecommand \EOS [0]{\spacefactor3000\relax}%
\providecommand \BibitemShut  [1]{\csname bibitem#1\endcsname}%
\let\auto@bib@innerbib\@empty
\bibitem [{\citenamefont {Vellekoop}\ and\ \citenamefont {Mosk}(2007)}]{vellekoop2007focusing}%
  \BibitemOpen
  \bibfield  {author} {\bibinfo {author} {\bibfnamefont {I.~M.}\ \bibnamefont {Vellekoop}}\ and\ \bibinfo {author} {\bibfnamefont {A.~P.}\ \bibnamefont {Mosk}},\ }\bibfield  {title} {\enquote {\bibinfo {title} {Focusing coherent light through opaque strongly scattering media},}\ }\href@noop {} {\bibfield  {journal} {\bibinfo  {journal} {Optics letters}\ }\textbf {\bibinfo {volume} {32}},\ \bibinfo {pages} {2309--2311} (\bibinfo {year} {2007})}\BibitemShut {NoStop}%
\bibitem [{\citenamefont {Mosk}\ \emph {et~al.}(2012)\citenamefont {Mosk}, \citenamefont {Lagendijk}, \citenamefont {Lerosey},\ and\ \citenamefont {Fink}}]{mosk2012controlling}%
  \BibitemOpen
  \bibfield  {author} {\bibinfo {author} {\bibfnamefont {A.~P.}\ \bibnamefont {Mosk}}, \bibinfo {author} {\bibfnamefont {A.}~\bibnamefont {Lagendijk}}, \bibinfo {author} {\bibfnamefont {G.}~\bibnamefont {Lerosey}}, \ and\ \bibinfo {author} {\bibfnamefont {M.}~\bibnamefont {Fink}},\ }\bibfield  {title} {\enquote {\bibinfo {title} {Controlling waves in space and time for imaging and focusing in complex media},}\ }\href@noop {} {\bibfield  {journal} {\bibinfo  {journal} {Nature photonics}\ }\textbf {\bibinfo {volume} {6}},\ \bibinfo {pages} {283--292} (\bibinfo {year} {2012})}\BibitemShut {NoStop}%
\bibitem [{\citenamefont {Popoff}\ \emph {et~al.}(2010)\citenamefont {Popoff}, \citenamefont {Lerosey}, \citenamefont {Carminati}, \citenamefont {Fink}, \citenamefont {Boccara},\ and\ \citenamefont {Gigan}}]{popoff2010measuring}%
  \BibitemOpen
  \bibfield  {author} {\bibinfo {author} {\bibfnamefont {S.~M.}\ \bibnamefont {Popoff}}, \bibinfo {author} {\bibfnamefont {G.}~\bibnamefont {Lerosey}}, \bibinfo {author} {\bibfnamefont {R.}~\bibnamefont {Carminati}}, \bibinfo {author} {\bibfnamefont {M.}~\bibnamefont {Fink}}, \bibinfo {author} {\bibfnamefont {A.~C.}\ \bibnamefont {Boccara}}, \ and\ \bibinfo {author} {\bibfnamefont {S.}~\bibnamefont {Gigan}},\ }\bibfield  {title} {\enquote {\bibinfo {title} {Measuring the transmission matrix in optics: an approach to the study and control of light propagation in disordered media},}\ }\href@noop {} {\bibfield  {journal} {\bibinfo  {journal} {Physical review letters}\ }\textbf {\bibinfo {volume} {104}},\ \bibinfo {pages} {100601} (\bibinfo {year} {2010})}\BibitemShut {NoStop}%
\bibitem [{\citenamefont {Vellekoop}(2015)}]{vellekoop2015feedback}%
  \BibitemOpen
  \bibfield  {author} {\bibinfo {author} {\bibfnamefont {I.~M.}\ \bibnamefont {Vellekoop}},\ }\bibfield  {title} {\enquote {\bibinfo {title} {Feedback-based wavefront shaping},}\ }\href@noop {} {\bibfield  {journal} {\bibinfo  {journal} {Optics express}\ }\textbf {\bibinfo {volume} {23}},\ \bibinfo {pages} {12189--12206} (\bibinfo {year} {2015})}\BibitemShut {NoStop}%
\bibitem [{\citenamefont {Cao}, \citenamefont {Mosk},\ and\ \citenamefont {Rotter}(2022)}]{cao2022shaping}%
  \BibitemOpen
  \bibfield  {author} {\bibinfo {author} {\bibfnamefont {H.}~\bibnamefont {Cao}}, \bibinfo {author} {\bibfnamefont {A.~P.}\ \bibnamefont {Mosk}}, \ and\ \bibinfo {author} {\bibfnamefont {S.}~\bibnamefont {Rotter}},\ }\bibfield  {title} {\enquote {\bibinfo {title} {Shaping the propagation of light in complex media},}\ }\href@noop {} {\bibfield  {journal} {\bibinfo  {journal} {Nature Physics}\ }\textbf {\bibinfo {volume} {18}},\ \bibinfo {pages} {994--1007} (\bibinfo {year} {2022})}\BibitemShut {NoStop}%
\bibitem [{\citenamefont {Gigan}\ \emph {et~al.}(2022)\citenamefont {Gigan}, \citenamefont {Katz}, \citenamefont {De~Aguiar}, \citenamefont {Andresen}, \citenamefont {Aubry}, \citenamefont {Bertolotti}, \citenamefont {Bossy}, \citenamefont {Bouchet}, \citenamefont {Brake}, \citenamefont {Brasselet} \emph {et~al.}}]{gigan2022roadmap}%
  \BibitemOpen
  \bibfield  {author} {\bibinfo {author} {\bibfnamefont {S.}~\bibnamefont {Gigan}}, \bibinfo {author} {\bibfnamefont {O.}~\bibnamefont {Katz}}, \bibinfo {author} {\bibfnamefont {H.~B.}\ \bibnamefont {De~Aguiar}}, \bibinfo {author} {\bibfnamefont {E.~R.}\ \bibnamefont {Andresen}}, \bibinfo {author} {\bibfnamefont {A.}~\bibnamefont {Aubry}}, \bibinfo {author} {\bibfnamefont {J.}~\bibnamefont {Bertolotti}}, \bibinfo {author} {\bibfnamefont {E.}~\bibnamefont {Bossy}}, \bibinfo {author} {\bibfnamefont {D.}~\bibnamefont {Bouchet}}, \bibinfo {author} {\bibfnamefont {J.}~\bibnamefont {Brake}}, \bibinfo {author} {\bibfnamefont {S.}~\bibnamefont {Brasselet}},  \emph {et~al.},\ }\bibfield  {title} {\enquote {\bibinfo {title} {Roadmap on wavefront shaping and deep imaging in complex media},}\ }\href@noop {} {\bibfield  {journal} {\bibinfo  {journal} {Journal of Physics: Photonics}\ }\textbf {\bibinfo {volume} {4}},\ \bibinfo {pages} {042501} (\bibinfo {year} {2022})}\BibitemShut {NoStop}%
\bibitem [{\citenamefont {Lib}\ and\ \citenamefont {Bromberg}(2022{\natexlab{a}})}]{lib2022quantum}%
  \BibitemOpen
  \bibfield  {author} {\bibinfo {author} {\bibfnamefont {O.}~\bibnamefont {Lib}}\ and\ \bibinfo {author} {\bibfnamefont {Y.}~\bibnamefont {Bromberg}},\ }\bibfield  {title} {\enquote {\bibinfo {title} {Quantum light in complex media and its applications},}\ }\href@noop {} {\bibfield  {journal} {\bibinfo  {journal} {Nature Physics}\ }\textbf {\bibinfo {volume} {18}},\ \bibinfo {pages} {986--993} (\bibinfo {year} {2022}{\natexlab{a}})}\BibitemShut {NoStop}%
\bibitem [{\citenamefont {Goyal}\ \emph {et~al.}(2016)\citenamefont {Goyal}, \citenamefont {Roux}, \citenamefont {Konrad}, \citenamefont {Forbes} \emph {et~al.}}]{goyal2016effect}%
  \BibitemOpen
  \bibfield  {author} {\bibinfo {author} {\bibfnamefont {S.~K.}\ \bibnamefont {Goyal}}, \bibinfo {author} {\bibfnamefont {F.~S.}\ \bibnamefont {Roux}}, \bibinfo {author} {\bibfnamefont {T.}~\bibnamefont {Konrad}}, \bibinfo {author} {\bibfnamefont {A.}~\bibnamefont {Forbes}},  \emph {et~al.},\ }\bibfield  {title} {\enquote {\bibinfo {title} {The effect of turbulence on entanglement-based free-space quantum key distribution with photonic orbital angular momentum},}\ }\href@noop {} {\bibfield  {journal} {\bibinfo  {journal} {Journal of Optics}\ }\textbf {\bibinfo {volume} {18}},\ \bibinfo {pages} {064002} (\bibinfo {year} {2016})}\BibitemShut {NoStop}%
\bibitem [{\citenamefont {Ndagano}\ \emph {et~al.}(2017)\citenamefont {Ndagano}, \citenamefont {Perez-Garcia}, \citenamefont {Roux}, \citenamefont {McLaren}, \citenamefont {Rosales-Guzman}, \citenamefont {Zhang}, \citenamefont {Mouane}, \citenamefont {Hernandez-Aranda}, \citenamefont {Konrad},\ and\ \citenamefont {Forbes}}]{ndagano2017characterizing}%
  \BibitemOpen
  \bibfield  {author} {\bibinfo {author} {\bibfnamefont {B.}~\bibnamefont {Ndagano}}, \bibinfo {author} {\bibfnamefont {B.}~\bibnamefont {Perez-Garcia}}, \bibinfo {author} {\bibfnamefont {F.~S.}\ \bibnamefont {Roux}}, \bibinfo {author} {\bibfnamefont {M.}~\bibnamefont {McLaren}}, \bibinfo {author} {\bibfnamefont {C.}~\bibnamefont {Rosales-Guzman}}, \bibinfo {author} {\bibfnamefont {Y.}~\bibnamefont {Zhang}}, \bibinfo {author} {\bibfnamefont {O.}~\bibnamefont {Mouane}}, \bibinfo {author} {\bibfnamefont {R.~I.}\ \bibnamefont {Hernandez-Aranda}}, \bibinfo {author} {\bibfnamefont {T.}~\bibnamefont {Konrad}}, \ and\ \bibinfo {author} {\bibfnamefont {A.}~\bibnamefont {Forbes}},\ }\bibfield  {title} {\enquote {\bibinfo {title} {Characterizing quantum channels with non-separable states of classical light},}\ }\href@noop {} {\bibfield  {journal} {\bibinfo  {journal} {Nature Physics}\ }\textbf {\bibinfo {volume} {13}},\ \bibinfo {pages} {397--402} (\bibinfo {year} {2017})}\BibitemShut {NoStop}%
\bibitem [{\citenamefont {Liu}\ \emph {et~al.}(2019)\citenamefont {Liu}, \citenamefont {Pang}, \citenamefont {Zhao}, \citenamefont {Liao}, \citenamefont {Zhang}, \citenamefont {Song}, \citenamefont {Cao}, \citenamefont {Du}, \citenamefont {Li}, \citenamefont {Song} \emph {et~al.}}]{liu2019single}%
  \BibitemOpen
  \bibfield  {author} {\bibinfo {author} {\bibfnamefont {C.}~\bibnamefont {Liu}}, \bibinfo {author} {\bibfnamefont {K.}~\bibnamefont {Pang}}, \bibinfo {author} {\bibfnamefont {Z.}~\bibnamefont {Zhao}}, \bibinfo {author} {\bibfnamefont {P.}~\bibnamefont {Liao}}, \bibinfo {author} {\bibfnamefont {R.}~\bibnamefont {Zhang}}, \bibinfo {author} {\bibfnamefont {H.}~\bibnamefont {Song}}, \bibinfo {author} {\bibfnamefont {Y.}~\bibnamefont {Cao}}, \bibinfo {author} {\bibfnamefont {J.}~\bibnamefont {Du}}, \bibinfo {author} {\bibfnamefont {L.}~\bibnamefont {Li}}, \bibinfo {author} {\bibfnamefont {H.}~\bibnamefont {Song}},  \emph {et~al.},\ }\bibfield  {title} {\enquote {\bibinfo {title} {Single-end adaptive optics compensation for emulated turbulence in a bi-directional 10-mbit/s per channel free-space quantum communication link using orbital-angular-momentum encoding},}\ }\href@noop {} {\bibfield  {journal} {\bibinfo  {journal} {Research}\ } (\bibinfo {year} {2019})}\BibitemShut {NoStop}%
\bibitem [{\citenamefont {Cao}\ \emph {et~al.}(2020)\citenamefont {Cao}, \citenamefont {Li}, \citenamefont {Yang}, \citenamefont {Jiang}, \citenamefont {Li}, \citenamefont {Hu}, \citenamefont {Abulizi}, \citenamefont {Li}, \citenamefont {Zhang}, \citenamefont {Sun} \emph {et~al.}}]{cao2020long}%
  \BibitemOpen
  \bibfield  {author} {\bibinfo {author} {\bibfnamefont {Y.}~\bibnamefont {Cao}}, \bibinfo {author} {\bibfnamefont {Y.-H.}\ \bibnamefont {Li}}, \bibinfo {author} {\bibfnamefont {K.-X.}\ \bibnamefont {Yang}}, \bibinfo {author} {\bibfnamefont {Y.-F.}\ \bibnamefont {Jiang}}, \bibinfo {author} {\bibfnamefont {S.-L.}\ \bibnamefont {Li}}, \bibinfo {author} {\bibfnamefont {X.-L.}\ \bibnamefont {Hu}}, \bibinfo {author} {\bibfnamefont {M.}~\bibnamefont {Abulizi}}, \bibinfo {author} {\bibfnamefont {C.-L.}\ \bibnamefont {Li}}, \bibinfo {author} {\bibfnamefont {W.}~\bibnamefont {Zhang}}, \bibinfo {author} {\bibfnamefont {Q.-C.}\ \bibnamefont {Sun}},  \emph {et~al.},\ }\bibfield  {title} {\enquote {\bibinfo {title} {Long-distance free-space measurement-device-independent quantum key distribution},}\ }\href@noop {} {\bibfield  {journal} {\bibinfo  {journal} {Physical Review Letters}\ }\textbf {\bibinfo {volume} {125}},\ \bibinfo {pages} {260503} (\bibinfo {year} {2020})}\BibitemShut {NoStop}%
\bibitem [{\citenamefont {Amitonova}\ \emph {et~al.}(2020)\citenamefont {Amitonova}, \citenamefont {Tentrup}, \citenamefont {Vellekoop},\ and\ \citenamefont {Pinkse}}]{amitonova2020quantum}%
  \BibitemOpen
  \bibfield  {author} {\bibinfo {author} {\bibfnamefont {L.~V.}\ \bibnamefont {Amitonova}}, \bibinfo {author} {\bibfnamefont {T.~B.}\ \bibnamefont {Tentrup}}, \bibinfo {author} {\bibfnamefont {I.~M.}\ \bibnamefont {Vellekoop}}, \ and\ \bibinfo {author} {\bibfnamefont {P.~W.}\ \bibnamefont {Pinkse}},\ }\bibfield  {title} {\enquote {\bibinfo {title} {Quantum key establishment via a multimode fiber},}\ }\href@noop {} {\bibfield  {journal} {\bibinfo  {journal} {Optics express}\ }\textbf {\bibinfo {volume} {28}},\ \bibinfo {pages} {5965--5981} (\bibinfo {year} {2020})}\BibitemShut {NoStop}%
\bibitem [{\citenamefont {Pugh}\ \emph {et~al.}(2020)\citenamefont {Pugh}, \citenamefont {Lavigne}, \citenamefont {Bourgoin}, \citenamefont {Higgins},\ and\ \citenamefont {Jennewein}}]{pugh2020adaptive}%
  \BibitemOpen
  \bibfield  {author} {\bibinfo {author} {\bibfnamefont {C.~J.}\ \bibnamefont {Pugh}}, \bibinfo {author} {\bibfnamefont {J.-F.}\ \bibnamefont {Lavigne}}, \bibinfo {author} {\bibfnamefont {J.-P.}\ \bibnamefont {Bourgoin}}, \bibinfo {author} {\bibfnamefont {B.~L.}\ \bibnamefont {Higgins}}, \ and\ \bibinfo {author} {\bibfnamefont {T.}~\bibnamefont {Jennewein}},\ }\bibfield  {title} {\enquote {\bibinfo {title} {Adaptive optics benefit for quantum key distribution uplink from ground to a satellite},}\ }\href@noop {} {\bibfield  {journal} {\bibinfo  {journal} {Advanced Optical Technologies}\ }\textbf {\bibinfo {volume} {9}},\ \bibinfo {pages} {263--273} (\bibinfo {year} {2020})}\BibitemShut {NoStop}%
\bibitem [{\citenamefont {Beenakker}, \citenamefont {Venderbos},\ and\ \citenamefont {Van~Exter}(2009)}]{beenakker2009two}%
  \BibitemOpen
  \bibfield  {author} {\bibinfo {author} {\bibfnamefont {C.}~\bibnamefont {Beenakker}}, \bibinfo {author} {\bibfnamefont {J.}~\bibnamefont {Venderbos}}, \ and\ \bibinfo {author} {\bibfnamefont {M.}~\bibnamefont {Van~Exter}},\ }\bibfield  {title} {\enquote {\bibinfo {title} {Two-photon speckle as a probe of multi-dimensional entanglement},}\ }\href@noop {} {\bibfield  {journal} {\bibinfo  {journal} {Physical review letters}\ }\textbf {\bibinfo {volume} {102}},\ \bibinfo {pages} {193601} (\bibinfo {year} {2009})}\BibitemShut {NoStop}%
\bibitem [{\citenamefont {Peeters}, \citenamefont {Moerman},\ and\ \citenamefont {Van~Exter}(2010)}]{peeters2010observation}%
  \BibitemOpen
  \bibfield  {author} {\bibinfo {author} {\bibfnamefont {W.}~\bibnamefont {Peeters}}, \bibinfo {author} {\bibfnamefont {J.}~\bibnamefont {Moerman}}, \ and\ \bibinfo {author} {\bibfnamefont {M.}~\bibnamefont {Van~Exter}},\ }\bibfield  {title} {\enquote {\bibinfo {title} {Observation of two-photon speckle patterns},}\ }\href@noop {} {\bibfield  {journal} {\bibinfo  {journal} {Physical review letters}\ }\textbf {\bibinfo {volume} {104}},\ \bibinfo {pages} {173601} (\bibinfo {year} {2010})}\BibitemShut {NoStop}%
\bibitem [{\citenamefont {Carpenter}\ \emph {et~al.}(2013)\citenamefont {Carpenter}, \citenamefont {Xiong}, \citenamefont {Collins}, \citenamefont {Li}, \citenamefont {Krauss}, \citenamefont {Eggleton}, \citenamefont {Clark},\ and\ \citenamefont {Schr{\"o}der}}]{carpenter2013mode}%
  \BibitemOpen
  \bibfield  {author} {\bibinfo {author} {\bibfnamefont {J.}~\bibnamefont {Carpenter}}, \bibinfo {author} {\bibfnamefont {C.}~\bibnamefont {Xiong}}, \bibinfo {author} {\bibfnamefont {M.~J.}\ \bibnamefont {Collins}}, \bibinfo {author} {\bibfnamefont {J.}~\bibnamefont {Li}}, \bibinfo {author} {\bibfnamefont {T.~F.}\ \bibnamefont {Krauss}}, \bibinfo {author} {\bibfnamefont {B.~J.}\ \bibnamefont {Eggleton}}, \bibinfo {author} {\bibfnamefont {A.~S.}\ \bibnamefont {Clark}}, \ and\ \bibinfo {author} {\bibfnamefont {J.}~\bibnamefont {Schr{\"o}der}},\ }\bibfield  {title} {\enquote {\bibinfo {title} {Mode multiplexed single-photon and classical channels in a few-mode fiber},}\ }\href@noop {} {\bibfield  {journal} {\bibinfo  {journal} {Optics express}\ }\textbf {\bibinfo {volume} {21}},\ \bibinfo {pages} {28794--28800} (\bibinfo {year} {2013})}\BibitemShut {NoStop}%
\bibitem [{\citenamefont {Defienne}\ \emph {et~al.}(2014)\citenamefont {Defienne}, \citenamefont {Barbieri}, \citenamefont {Chalopin}, \citenamefont {Chatel}, \citenamefont {Walmsley}, \citenamefont {Smith},\ and\ \citenamefont {Gigan}}]{defienne2014nonclassical}%
  \BibitemOpen
  \bibfield  {author} {\bibinfo {author} {\bibfnamefont {H.}~\bibnamefont {Defienne}}, \bibinfo {author} {\bibfnamefont {M.}~\bibnamefont {Barbieri}}, \bibinfo {author} {\bibfnamefont {B.}~\bibnamefont {Chalopin}}, \bibinfo {author} {\bibfnamefont {B.}~\bibnamefont {Chatel}}, \bibinfo {author} {\bibfnamefont {I.}~\bibnamefont {Walmsley}}, \bibinfo {author} {\bibfnamefont {B.}~\bibnamefont {Smith}}, \ and\ \bibinfo {author} {\bibfnamefont {S.}~\bibnamefont {Gigan}},\ }\bibfield  {title} {\enquote {\bibinfo {title} {Nonclassical light manipulation in a multiple-scattering medium},}\ }\href@noop {} {\bibfield  {journal} {\bibinfo  {journal} {Optics letters}\ }\textbf {\bibinfo {volume} {39}},\ \bibinfo {pages} {6090--6093} (\bibinfo {year} {2014})}\BibitemShut {NoStop}%
\bibitem [{\citenamefont {Huisman}\ \emph {et~al.}(2014)\citenamefont {Huisman}, \citenamefont {Huisman}, \citenamefont {Mosk},\ and\ \citenamefont {Pinkse}}]{huisman2014controlling}%
  \BibitemOpen
  \bibfield  {author} {\bibinfo {author} {\bibfnamefont {T.~J.}\ \bibnamefont {Huisman}}, \bibinfo {author} {\bibfnamefont {S.~R.}\ \bibnamefont {Huisman}}, \bibinfo {author} {\bibfnamefont {A.~P.}\ \bibnamefont {Mosk}}, \ and\ \bibinfo {author} {\bibfnamefont {P.~W.}\ \bibnamefont {Pinkse}},\ }\bibfield  {title} {\enquote {\bibinfo {title} {Controlling single-photon fock-state propagation through opaque scattering media},}\ }\href@noop {} {\bibfield  {journal} {\bibinfo  {journal} {Applied Physics B}\ }\textbf {\bibinfo {volume} {116}},\ \bibinfo {pages} {603--607} (\bibinfo {year} {2014})}\BibitemShut {NoStop}%
\bibitem [{\citenamefont {Valencia}\ \emph {et~al.}(2020)\citenamefont {Valencia}, \citenamefont {Goel}, \citenamefont {McCutcheon}, \citenamefont {Defienne},\ and\ \citenamefont {Malik}}]{valencia2020unscrambling}%
  \BibitemOpen
  \bibfield  {author} {\bibinfo {author} {\bibfnamefont {N.~H.}\ \bibnamefont {Valencia}}, \bibinfo {author} {\bibfnamefont {S.}~\bibnamefont {Goel}}, \bibinfo {author} {\bibfnamefont {W.}~\bibnamefont {McCutcheon}}, \bibinfo {author} {\bibfnamefont {H.}~\bibnamefont {Defienne}}, \ and\ \bibinfo {author} {\bibfnamefont {M.}~\bibnamefont {Malik}},\ }\bibfield  {title} {\enquote {\bibinfo {title} {Unscrambling entanglement through a complex medium},}\ }\href@noop {} {\bibfield  {journal} {\bibinfo  {journal} {Nature Physics}\ }\textbf {\bibinfo {volume} {16}},\ \bibinfo {pages} {1112--1116} (\bibinfo {year} {2020})}\BibitemShut {NoStop}%
\bibitem [{\citenamefont {Shekel}\ \emph {et~al.}(2023)\citenamefont {Shekel}, \citenamefont {Lib}, \citenamefont {Gutiérrez-Cuevas}, \citenamefont {Popoff}, \citenamefont {Ling},\ and\ \citenamefont {Bromberg}}]{shekel2023pianoq}%
  \BibitemOpen
  \bibfield  {author} {\bibinfo {author} {\bibfnamefont {R.}~\bibnamefont {Shekel}}, \bibinfo {author} {\bibfnamefont {O.}~\bibnamefont {Lib}}, \bibinfo {author} {\bibfnamefont {R.}~\bibnamefont {Gutiérrez-Cuevas}}, \bibinfo {author} {\bibfnamefont {S.~M.}\ \bibnamefont {Popoff}}, \bibinfo {author} {\bibfnamefont {A.}~\bibnamefont {Ling}}, \ and\ \bibinfo {author} {\bibfnamefont {Y.}~\bibnamefont {Bromberg}},\ }\bibfield  {title} {\enquote {\bibinfo {title} {{Shaping single photons through multimode optical fibers using mechanical perturbations}},}\ }\href@noop {} {\bibfield  {journal} {\bibinfo  {journal} {APL Photonics}\ }\textbf {\bibinfo {volume} {8}},\ \bibinfo {pages} {096109} (\bibinfo {year} {2023})}\BibitemShut {NoStop}%
\bibitem [{\citenamefont {Devaux}\ \emph {et~al.}(2023)\citenamefont {Devaux}, \citenamefont {Mosset}, \citenamefont {Popoff},\ and\ \citenamefont {Lantz}}]{devaux2023restoring}%
  \BibitemOpen
  \bibfield  {author} {\bibinfo {author} {\bibfnamefont {F.}~\bibnamefont {Devaux}}, \bibinfo {author} {\bibfnamefont {A.}~\bibnamefont {Mosset}}, \bibinfo {author} {\bibfnamefont {S.~M.}\ \bibnamefont {Popoff}}, \ and\ \bibinfo {author} {\bibfnamefont {E.}~\bibnamefont {Lantz}},\ }\bibfield  {title} {\enquote {\bibinfo {title} {Restoring and tailoring very high dimensional spatial entanglement of a biphoton state transmitted through a scattering medium},}\ }\href@noop {} {\bibfield  {journal} {\bibinfo  {journal} {Journal of Optics}\ }\textbf {\bibinfo {volume} {25}},\ \bibinfo {pages} {055201} (\bibinfo {year} {2023})}\BibitemShut {NoStop}%
\bibitem [{\citenamefont {Defienne}\ \emph {et~al.}(2016)\citenamefont {Defienne}, \citenamefont {Barbieri}, \citenamefont {Walmsley}, \citenamefont {Smith},\ and\ \citenamefont {Gigan}}]{defienne2016two}%
  \BibitemOpen
  \bibfield  {author} {\bibinfo {author} {\bibfnamefont {H.}~\bibnamefont {Defienne}}, \bibinfo {author} {\bibfnamefont {M.}~\bibnamefont {Barbieri}}, \bibinfo {author} {\bibfnamefont {I.~A.}\ \bibnamefont {Walmsley}}, \bibinfo {author} {\bibfnamefont {B.~J.}\ \bibnamefont {Smith}}, \ and\ \bibinfo {author} {\bibfnamefont {S.}~\bibnamefont {Gigan}},\ }\bibfield  {title} {\enquote {\bibinfo {title} {Two-photon quantum walk in a multimode fiber},}\ }\href@noop {} {\bibfield  {journal} {\bibinfo  {journal} {Science advances}\ }\textbf {\bibinfo {volume} {2}},\ \bibinfo {pages} {e1501054} (\bibinfo {year} {2016})}\BibitemShut {NoStop}%
\bibitem [{\citenamefont {Wolterink}\ \emph {et~al.}(2016)\citenamefont {Wolterink}, \citenamefont {Uppu}, \citenamefont {Ctistis}, \citenamefont {Vos}, \citenamefont {Boller},\ and\ \citenamefont {Pinkse}}]{pinkse2016programmable}%
  \BibitemOpen
  \bibfield  {author} {\bibinfo {author} {\bibfnamefont {T.~A.~W.}\ \bibnamefont {Wolterink}}, \bibinfo {author} {\bibfnamefont {R.}~\bibnamefont {Uppu}}, \bibinfo {author} {\bibfnamefont {G.}~\bibnamefont {Ctistis}}, \bibinfo {author} {\bibfnamefont {W.~L.}\ \bibnamefont {Vos}}, \bibinfo {author} {\bibfnamefont {K.-J.}\ \bibnamefont {Boller}}, \ and\ \bibinfo {author} {\bibfnamefont {P.~W.~H.}\ \bibnamefont {Pinkse}},\ }\bibfield  {title} {\enquote {\bibinfo {title} {Programmable two-photon quantum interference in ${10}^{3}$ channels in opaque scattering media},}\ }\href@noop {} {\bibfield  {journal} {\bibinfo  {journal} {Phys. Rev. A}\ }\textbf {\bibinfo {volume} {93}},\ \bibinfo {pages} {053817} (\bibinfo {year} {2016})}\BibitemShut {NoStop}%
\bibitem [{\citenamefont {Defienne}, \citenamefont {Reichert},\ and\ \citenamefont {Fleischer}(2018)}]{defienne2018adaptive}%
  \BibitemOpen
  \bibfield  {author} {\bibinfo {author} {\bibfnamefont {H.}~\bibnamefont {Defienne}}, \bibinfo {author} {\bibfnamefont {M.}~\bibnamefont {Reichert}}, \ and\ \bibinfo {author} {\bibfnamefont {J.~W.}\ \bibnamefont {Fleischer}},\ }\bibfield  {title} {\enquote {\bibinfo {title} {Adaptive quantum optics with spatially entangled photon pairs},}\ }\href@noop {} {\bibfield  {journal} {\bibinfo  {journal} {Physical review letters}\ }\textbf {\bibinfo {volume} {121}},\ \bibinfo {pages} {233601} (\bibinfo {year} {2018})}\BibitemShut {NoStop}%
\bibitem [{\citenamefont {Peng}\ \emph {et~al.}(2018)\citenamefont {Peng}, \citenamefont {Qiao}, \citenamefont {Xiang},\ and\ \citenamefont {Chen}}]{Peng2018manipulation}%
  \BibitemOpen
  \bibfield  {author} {\bibinfo {author} {\bibfnamefont {Y.}~\bibnamefont {Peng}}, \bibinfo {author} {\bibfnamefont {Y.}~\bibnamefont {Qiao}}, \bibinfo {author} {\bibfnamefont {T.}~\bibnamefont {Xiang}}, \ and\ \bibinfo {author} {\bibfnamefont {X.}~\bibnamefont {Chen}},\ }\bibfield  {title} {\enquote {\bibinfo {title} {Manipulation of the spontaneous parametric down-conversion process in space and frequency domains via wavefront shaping},}\ }\href@noop {} {\bibfield  {journal} {\bibinfo  {journal} {Opt. Lett.}\ }\textbf {\bibinfo {volume} {43}},\ \bibinfo {pages} {3985--3988} (\bibinfo {year} {2018})}\BibitemShut {NoStop}%
\bibitem [{\citenamefont {Lib}, \citenamefont {Hasson},\ and\ \citenamefont {Bromberg}(2020)}]{lib2020real}%
  \BibitemOpen
  \bibfield  {author} {\bibinfo {author} {\bibfnamefont {O.}~\bibnamefont {Lib}}, \bibinfo {author} {\bibfnamefont {G.}~\bibnamefont {Hasson}}, \ and\ \bibinfo {author} {\bibfnamefont {Y.}~\bibnamefont {Bromberg}},\ }\bibfield  {title} {\enquote {\bibinfo {title} {Real-time shaping of entangled photons by classical control and feedback},}\ }\href@noop {} {\bibfield  {journal} {\bibinfo  {journal} {Science Advances}\ }\textbf {\bibinfo {volume} {6}},\ \bibinfo {pages} {eabb6298} (\bibinfo {year} {2020})}\BibitemShut {NoStop}%
\bibitem [{\citenamefont {Lib}\ and\ \citenamefont {Bromberg}(2020)}]{lib2020pump}%
  \BibitemOpen
  \bibfield  {author} {\bibinfo {author} {\bibfnamefont {O.}~\bibnamefont {Lib}}\ and\ \bibinfo {author} {\bibfnamefont {Y.}~\bibnamefont {Bromberg}},\ }\bibfield  {title} {\enquote {\bibinfo {title} {Pump-shaping of non-collinear and non-degenerate entangled photons},}\ }\href@noop {} {\bibfield  {journal} {\bibinfo  {journal} {Optics Letters}\ }\textbf {\bibinfo {volume} {45}},\ \bibinfo {pages} {6827--6830} (\bibinfo {year} {2020})}\BibitemShut {NoStop}%
\bibitem [{\citenamefont {Soro}\ \emph {et~al.}(2021)\citenamefont {Soro}, \citenamefont {Lantz}, \citenamefont {Mosset},\ and\ \citenamefont {Devaux}}]{soro2021quantum}%
  \BibitemOpen
  \bibfield  {author} {\bibinfo {author} {\bibfnamefont {G.}~\bibnamefont {Soro}}, \bibinfo {author} {\bibfnamefont {E.}~\bibnamefont {Lantz}}, \bibinfo {author} {\bibfnamefont {A.}~\bibnamefont {Mosset}}, \ and\ \bibinfo {author} {\bibfnamefont {F.}~\bibnamefont {Devaux}},\ }\bibfield  {title} {\enquote {\bibinfo {title} {Quantum spatial correlations imaging through thick scattering media: experiments and comparison with simulations of the biphoton wave function},}\ }\href@noop {} {\bibfield  {journal} {\bibinfo  {journal} {Journal of Optics}\ }\textbf {\bibinfo {volume} {23}},\ \bibinfo {pages} {025201} (\bibinfo {year} {2021})}\BibitemShut {NoStop}%
\bibitem [{\citenamefont {Cameron}\ \emph {et~al.}(2024)\citenamefont {Cameron}, \citenamefont {Courme}, \citenamefont {Vernière}, \citenamefont {Pandya}, \citenamefont {Faccio},\ and\ \citenamefont {Defienne}}]{cameron2024adaptive}%
  \BibitemOpen
  \bibfield  {author} {\bibinfo {author} {\bibfnamefont {P.}~\bibnamefont {Cameron}}, \bibinfo {author} {\bibfnamefont {B.}~\bibnamefont {Courme}}, \bibinfo {author} {\bibfnamefont {C.}~\bibnamefont {Vernière}}, \bibinfo {author} {\bibfnamefont {R.}~\bibnamefont {Pandya}}, \bibinfo {author} {\bibfnamefont {D.}~\bibnamefont {Faccio}}, \ and\ \bibinfo {author} {\bibfnamefont {H.}~\bibnamefont {Defienne}},\ }\bibfield  {title} {\enquote {\bibinfo {title} {Adaptive optical imaging with entangled photons},}\ }\href@noop {} {\bibfield  {journal} {\bibinfo  {journal} {Science}\ }\textbf {\bibinfo {volume} {383}},\ \bibinfo {pages} {1142--1148} (\bibinfo {year} {2024})}\BibitemShut {NoStop}%
\bibitem [{\citenamefont {Shekel}, \citenamefont {Lib},\ and\ \citenamefont {Bromberg}(2024)}]{shekel2024shaping}%
  \BibitemOpen
  \bibfield  {author} {\bibinfo {author} {\bibfnamefont {R.}~\bibnamefont {Shekel}}, \bibinfo {author} {\bibfnamefont {O.}~\bibnamefont {Lib}}, \ and\ \bibinfo {author} {\bibfnamefont {Y.}~\bibnamefont {Bromberg}},\ }\bibfield  {title} {\enquote {\bibinfo {title} {Shaping entangled photons through arbitrary scattering media using an advanced wave beacon},}\ }\href@noop {} {\bibfield  {journal} {\bibinfo  {journal} {Optica Quantum}\ }\textbf {\bibinfo {volume} {2}},\ \bibinfo {pages} {303--309} (\bibinfo {year} {2024})}\BibitemShut {NoStop}%
\bibitem [{\citenamefont {Courme}\ \emph {et~al.}(2025)\citenamefont {Courme}, \citenamefont {Verni{\`e}re}, \citenamefont {Joly}, \citenamefont {Faccio}, \citenamefont {Gigan},\ and\ \citenamefont {Defienne}}]{courme2025non}%
  \BibitemOpen
  \bibfield  {author} {\bibinfo {author} {\bibfnamefont {B.}~\bibnamefont {Courme}}, \bibinfo {author} {\bibfnamefont {C.}~\bibnamefont {Verni{\`e}re}}, \bibinfo {author} {\bibfnamefont {M.}~\bibnamefont {Joly}}, \bibinfo {author} {\bibfnamefont {D.}~\bibnamefont {Faccio}}, \bibinfo {author} {\bibfnamefont {S.}~\bibnamefont {Gigan}}, \ and\ \bibinfo {author} {\bibfnamefont {H.}~\bibnamefont {Defienne}},\ }\bibfield  {title} {\enquote {\bibinfo {title} {Non-classical optimization through complex media},}\ }\href@noop {} {\bibfield  {journal} {\bibinfo  {journal} {arXiv preprint arXiv:2503.24283}\ } (\bibinfo {year} {2025})}\BibitemShut {NoStop}%
\bibitem [{\citenamefont {Shekel}\ \emph {et~al.}(2021)\citenamefont {Shekel}, \citenamefont {Lib}, \citenamefont {Sardas},\ and\ \citenamefont {Bromberg}}]{shekel2021shaping}%
  \BibitemOpen
  \bibfield  {author} {\bibinfo {author} {\bibfnamefont {R.}~\bibnamefont {Shekel}}, \bibinfo {author} {\bibfnamefont {O.}~\bibnamefont {Lib}}, \bibinfo {author} {\bibfnamefont {A.}~\bibnamefont {Sardas}}, \ and\ \bibinfo {author} {\bibfnamefont {Y.}~\bibnamefont {Bromberg}},\ }\bibfield  {title} {\enquote {\bibinfo {title} {Shaping entangled photons through emulated turbulent atmosphere},}\ }\href@noop {} {\bibfield  {journal} {\bibinfo  {journal} {OSA Continuum}\ }\textbf {\bibinfo {volume} {4}},\ \bibinfo {pages} {2339--2350} (\bibinfo {year} {2021})}\BibitemShut {NoStop}%
\bibitem [{\citenamefont {Vellekoop}\ and\ \citenamefont {Mosk}(2008)}]{vellekoop2008phase}%
  \BibitemOpen
  \bibfield  {author} {\bibinfo {author} {\bibfnamefont {I.~M.}\ \bibnamefont {Vellekoop}}\ and\ \bibinfo {author} {\bibfnamefont {A.}~\bibnamefont {Mosk}},\ }\bibfield  {title} {\enquote {\bibinfo {title} {Phase control algorithms for focusing light through turbid media},}\ }\href@noop {} {\bibfield  {journal} {\bibinfo  {journal} {Optics communications}\ }\textbf {\bibinfo {volume} {281}},\ \bibinfo {pages} {3071--3080} (\bibinfo {year} {2008})}\BibitemShut {NoStop}%
\bibitem [{\citenamefont {Walborn}\ \emph {et~al.}(2010)\citenamefont {Walborn}, \citenamefont {Monken}, \citenamefont {P{\'a}dua},\ and\ \citenamefont {Ribeiro}}]{walborn2010spatial}%
  \BibitemOpen
  \bibfield  {author} {\bibinfo {author} {\bibfnamefont {S.~P.}\ \bibnamefont {Walborn}}, \bibinfo {author} {\bibfnamefont {C.}~\bibnamefont {Monken}}, \bibinfo {author} {\bibfnamefont {S.}~\bibnamefont {P{\'a}dua}}, \ and\ \bibinfo {author} {\bibfnamefont {P.~S.}\ \bibnamefont {Ribeiro}},\ }\bibfield  {title} {\enquote {\bibinfo {title} {Spatial correlations in parametric down-conversion},}\ }\href@noop {} {\bibfield  {journal} {\bibinfo  {journal} {Physics Reports}\ }\textbf {\bibinfo {volume} {495}},\ \bibinfo {pages} {87--139} (\bibinfo {year} {2010})}\BibitemShut {NoStop}%
\bibitem [{\citenamefont {Howell}\ \emph {et~al.}(2004)\citenamefont {Howell}, \citenamefont {Bennink}, \citenamefont {Bentley},\ and\ \citenamefont {Boyd}}]{howell2004realization}%
  \BibitemOpen
  \bibfield  {author} {\bibinfo {author} {\bibfnamefont {J.~C.}\ \bibnamefont {Howell}}, \bibinfo {author} {\bibfnamefont {R.~S.}\ \bibnamefont {Bennink}}, \bibinfo {author} {\bibfnamefont {S.~J.}\ \bibnamefont {Bentley}}, \ and\ \bibinfo {author} {\bibfnamefont {R.~W.}\ \bibnamefont {Boyd}},\ }\bibfield  {title} {\enquote {\bibinfo {title} {Realization of the einstein-podolsky-rosen paradox using momentum- and position-entangled photons from spontaneous parametric down conversion},}\ }\href@noop {} {\bibfield  {journal} {\bibinfo  {journal} {Physical review letters}\ }\textbf {\bibinfo {volume} {92}},\ \bibinfo {pages} {210403} (\bibinfo {year} {2004})}\BibitemShut {NoStop}%
\bibitem [{\citenamefont {Goodman}(2007)}]{goodman2007speckle}%
  \BibitemOpen
  \bibfield  {author} {\bibinfo {author} {\bibfnamefont {J.~W.}\ \bibnamefont {Goodman}},\ }\href@noop {} {\emph {\bibinfo {title} {Speckle phenomena in optics: theory and applications}}}\ (\bibinfo  {publisher} {Roberts and Company Publishers},\ \bibinfo {year} {2007})\BibitemShut {NoStop}%
\bibitem [{\citenamefont {Goetschy}\ and\ \citenamefont {Stone}(2013)}]{goetschy2013filtering}%
  \BibitemOpen
  \bibfield  {author} {\bibinfo {author} {\bibfnamefont {A.}~\bibnamefont {Goetschy}}\ and\ \bibinfo {author} {\bibfnamefont {A.}~\bibnamefont {Stone}},\ }\bibfield  {title} {\enquote {\bibinfo {title} {Filtering random matrices: the effect of incomplete channel control in multiple scattering},}\ }\href@noop {} {\bibfield  {journal} {\bibinfo  {journal} {Physical review letters}\ }\textbf {\bibinfo {volume} {111}},\ \bibinfo {pages} {063901} (\bibinfo {year} {2013})}\BibitemShut {NoStop}%
\bibitem [{\citenamefont {Klyshko}(1988)}]{klyshko1988simple}%
  \BibitemOpen
  \bibfield  {author} {\bibinfo {author} {\bibfnamefont {D.}~\bibnamefont {Klyshko}},\ }\bibfield  {title} {\enquote {\bibinfo {title} {A simple method of preparing pure states of an optical field, of implementing the einstein--podolsky--rosen experiment, and of demonstrating the complementarity principle},}\ }\href@noop {} {\bibfield  {journal} {\bibinfo  {journal} {Soviet Physics Uspekhi}\ }\textbf {\bibinfo {volume} {31}},\ \bibinfo {pages} {74} (\bibinfo {year} {1988})}\BibitemShut {NoStop}%
\bibitem [{\citenamefont {Belinskii}\ and\ \citenamefont {Klyshko}(1994)}]{belinskii1994two}%
  \BibitemOpen
  \bibfield  {author} {\bibinfo {author} {\bibfnamefont {A.}~\bibnamefont {Belinskii}}\ and\ \bibinfo {author} {\bibfnamefont {D.}~\bibnamefont {Klyshko}},\ }\bibfield  {title} {\enquote {\bibinfo {title} {Two-photon optics: diffraction, holography, and transformation of two-dimensional signals},}\ }\href@noop {} {\bibfield  {journal} {\bibinfo  {journal} {Soviet Journal of Experimental and Theoretical Physics}\ }\textbf {\bibinfo {volume} {78}},\ \bibinfo {pages} {259--262} (\bibinfo {year} {1994})}\BibitemShut {NoStop}%
\bibitem [{\citenamefont {Zheng}\ \emph {et~al.}(2024)\citenamefont {Zheng}, \citenamefont {Xu}, \citenamefont {Li},\ and\ \citenamefont {Guo}}]{zheng2024AWP}%
  \BibitemOpen
  \bibfield  {author} {\bibinfo {author} {\bibfnamefont {Y.}~\bibnamefont {Zheng}}, \bibinfo {author} {\bibfnamefont {J.-S.}\ \bibnamefont {Xu}}, \bibinfo {author} {\bibfnamefont {C.-F.}\ \bibnamefont {Li}}, \ and\ \bibinfo {author} {\bibfnamefont {G.-C.}\ \bibnamefont {Guo}},\ }\href {https://arxiv.org/abs/2412.02088} {\enquote {\bibinfo {title} {Theory of monochromatic advanced-wave picture and applications in biphoton optics},}\ } (\bibinfo {year} {2024}),\ \Eprint {http://arxiv.org/abs/2412.02088} {arXiv:2412.02088 [quant-ph]} \BibitemShut {NoStop}%
\bibitem [{\citenamefont {Cui}\ and\ \citenamefont {Yang}(2010)}]{cui2010implementation}%
  \BibitemOpen
  \bibfield  {author} {\bibinfo {author} {\bibfnamefont {M.}~\bibnamefont {Cui}}\ and\ \bibinfo {author} {\bibfnamefont {C.}~\bibnamefont {Yang}},\ }\bibfield  {title} {\enquote {\bibinfo {title} {Implementation of a digital optical phase conjugation system and its application to study the robustness of turbidity suppression by phase conjugation},}\ }\href@noop {} {\bibfield  {journal} {\bibinfo  {journal} {Optics express}\ }\textbf {\bibinfo {volume} {18}},\ \bibinfo {pages} {3444--3455} (\bibinfo {year} {2010})}\BibitemShut {NoStop}%
\bibitem [{\citenamefont {Boyd}, \citenamefont {Gaeta},\ and\ \citenamefont {Giese}(2008)}]{boyd2008nonlinear}%
  \BibitemOpen
  \bibfield  {author} {\bibinfo {author} {\bibfnamefont {R.~W.}\ \bibnamefont {Boyd}}, \bibinfo {author} {\bibfnamefont {A.~L.}\ \bibnamefont {Gaeta}}, \ and\ \bibinfo {author} {\bibfnamefont {E.}~\bibnamefont {Giese}},\ }\bibfield  {title} {\enquote {\bibinfo {title} {Nonlinear optics},}\ }in\ \href@noop {} {\emph {\bibinfo {booktitle} {Springer Handbook of Atomic, Molecular, and Optical Physics}}}\ (\bibinfo  {publisher} {Springer},\ \bibinfo {year} {2008})\ pp.\ \bibinfo {pages} {1097--1110}\BibitemShut {NoStop}%
\bibitem [{\citenamefont {Arora}\ \emph {et~al.}(2005)\citenamefont {Arora}, \citenamefont {Berger}, \citenamefont {Elad}, \citenamefont {Kindler},\ and\ \citenamefont {Safra}}]{arora2005non}%
  \BibitemOpen
  \bibfield  {author} {\bibinfo {author} {\bibfnamefont {S.}~\bibnamefont {Arora}}, \bibinfo {author} {\bibfnamefont {E.}~\bibnamefont {Berger}}, \bibinfo {author} {\bibfnamefont {H.}~\bibnamefont {Elad}}, \bibinfo {author} {\bibfnamefont {G.}~\bibnamefont {Kindler}}, \ and\ \bibinfo {author} {\bibfnamefont {M.}~\bibnamefont {Safra}},\ }\bibfield  {title} {\enquote {\bibinfo {title} {On non-approximability for quadratic programs},}\ }in\ \href@noop {} {\emph {\bibinfo {booktitle} {46th Annual IEEE Symposium on Foundations of Computer Science (FOCS'05)}}}\ (\bibinfo {organization} {IEEE},\ \bibinfo {year} {2005})\ pp.\ \bibinfo {pages} {206--215}\BibitemShut {NoStop}%
\bibitem [{\citenamefont {Pardalos}\ and\ \citenamefont {Vavasis}(1991)}]{pardalos1991quadratic}%
  \BibitemOpen
  \bibfield  {author} {\bibinfo {author} {\bibfnamefont {P.~M.}\ \bibnamefont {Pardalos}}\ and\ \bibinfo {author} {\bibfnamefont {S.~A.}\ \bibnamefont {Vavasis}},\ }\bibfield  {title} {\enquote {\bibinfo {title} {Quadratic programming with one negative eigenvalue is np-hard},}\ }\href@noop {} {\bibfield  {journal} {\bibinfo  {journal} {Journal of Global optimization}\ }\textbf {\bibinfo {volume} {1}},\ \bibinfo {pages} {15--22} (\bibinfo {year} {1991})}\BibitemShut {NoStop}%
\bibitem [{\citenamefont {Lucas}(2014)}]{lucas2014ising}%
  \BibitemOpen
  \bibfield  {author} {\bibinfo {author} {\bibfnamefont {A.}~\bibnamefont {Lucas}},\ }\bibfield  {title} {\enquote {\bibinfo {title} {Ising formulations of many np problems},}\ }\href@noop {} {\bibfield  {journal} {\bibinfo  {journal} {Frontiers in physics}\ }\textbf {\bibinfo {volume} {2}},\ \bibinfo {pages} {5} (\bibinfo {year} {2014})}\BibitemShut {NoStop}%
\bibitem [{\citenamefont {Pierangeli}, \citenamefont {Marcucci},\ and\ \citenamefont {Conti}(2019)}]{pierangeli2019large}%
  \BibitemOpen
  \bibfield  {author} {\bibinfo {author} {\bibfnamefont {D.}~\bibnamefont {Pierangeli}}, \bibinfo {author} {\bibfnamefont {G.}~\bibnamefont {Marcucci}}, \ and\ \bibinfo {author} {\bibfnamefont {C.}~\bibnamefont {Conti}},\ }\bibfield  {title} {\enquote {\bibinfo {title} {Large-scale photonic ising machine by spatial light modulation},}\ }\href@noop {} {\bibfield  {journal} {\bibinfo  {journal} {Physical review letters}\ }\textbf {\bibinfo {volume} {122}},\ \bibinfo {pages} {213902} (\bibinfo {year} {2019})}\BibitemShut {NoStop}%
\bibitem [{\citenamefont {Pierangeli}\ \emph {et~al.}(2021)\citenamefont {Pierangeli}, \citenamefont {Rafayelyan}, \citenamefont {Conti},\ and\ \citenamefont {Gigan}}]{pierangeli2021scalable}%
  \BibitemOpen
  \bibfield  {author} {\bibinfo {author} {\bibfnamefont {D.}~\bibnamefont {Pierangeli}}, \bibinfo {author} {\bibfnamefont {M.}~\bibnamefont {Rafayelyan}}, \bibinfo {author} {\bibfnamefont {C.}~\bibnamefont {Conti}}, \ and\ \bibinfo {author} {\bibfnamefont {S.}~\bibnamefont {Gigan}},\ }\bibfield  {title} {\enquote {\bibinfo {title} {Scalable spin-glass optical simulator},}\ }\href@noop {} {\bibfield  {journal} {\bibinfo  {journal} {Physical Review Applied}\ }\textbf {\bibinfo {volume} {15}},\ \bibinfo {pages} {034087} (\bibinfo {year} {2021})}\BibitemShut {NoStop}%
\bibitem [{qwf()}]{qwfs2024}%
  \BibitemOpen
  \href@noop {} {}\bibinfo {howpublished} {\url{https://doi.org/10.5281/zenodo.15300939}},\ \bibinfo {note} {code and data for publication: "Fundamental bounds of Wavefront Shaping of Spatially Entangled Photons"}\BibitemShut {NoStop}%
\bibitem [{\citenamefont {Akkermans}\ and\ \citenamefont {Montambaux}(2007)}]{akkermans2007mesoscopic}%
  \BibitemOpen
  \bibfield  {author} {\bibinfo {author} {\bibfnamefont {E.}~\bibnamefont {Akkermans}}\ and\ \bibinfo {author} {\bibfnamefont {G.}~\bibnamefont {Montambaux}},\ }\href@noop {} {\emph {\bibinfo {title} {Mesoscopic physics of electrons and photons}}}\ (\bibinfo  {publisher} {Cambridge university press},\ \bibinfo {year} {2007})\BibitemShut {NoStop}%
\bibitem [{\citenamefont {Carminati}\ and\ \citenamefont {Schotland}(2021)}]{carminati2021principles}%
  \BibitemOpen
  \bibfield  {author} {\bibinfo {author} {\bibfnamefont {R.}~\bibnamefont {Carminati}}\ and\ \bibinfo {author} {\bibfnamefont {J.~C.}\ \bibnamefont {Schotland}},\ }\href@noop {} {\emph {\bibinfo {title} {Principles of scattering and transport of light}}}\ (\bibinfo  {publisher} {Cambridge University Press},\ \bibinfo {year} {2021})\BibitemShut {NoStop}%
\bibitem [{\citenamefont {Lib}\ and\ \citenamefont {Bromberg}(2022{\natexlab{b}})}]{lib2022thermal}%
  \BibitemOpen
  \bibfield  {author} {\bibinfo {author} {\bibfnamefont {O.}~\bibnamefont {Lib}}\ and\ \bibinfo {author} {\bibfnamefont {Y.}~\bibnamefont {Bromberg}},\ }\bibfield  {title} {\enquote {\bibinfo {title} {Thermal biphotons},}\ }\href@noop {} {\bibfield  {journal} {\bibinfo  {journal} {APL Photonics}\ }\textbf {\bibinfo {volume} {7}} (\bibinfo {year} {2022}{\natexlab{b}})}\BibitemShut {NoStop}%
\bibitem [{\citenamefont {Leedumrongwatthanakun}\ \emph {et~al.}(2020)\citenamefont {Leedumrongwatthanakun}, \citenamefont {Innocenti}, \citenamefont {Defienne}, \citenamefont {Juffmann}, \citenamefont {Ferraro}, \citenamefont {Paternostro},\ and\ \citenamefont {Gigan}}]{leedumrongwatthanakun2020programmable}%
  \BibitemOpen
  \bibfield  {author} {\bibinfo {author} {\bibfnamefont {S.}~\bibnamefont {Leedumrongwatthanakun}}, \bibinfo {author} {\bibfnamefont {L.}~\bibnamefont {Innocenti}}, \bibinfo {author} {\bibfnamefont {H.}~\bibnamefont {Defienne}}, \bibinfo {author} {\bibfnamefont {T.}~\bibnamefont {Juffmann}}, \bibinfo {author} {\bibfnamefont {A.}~\bibnamefont {Ferraro}}, \bibinfo {author} {\bibfnamefont {M.}~\bibnamefont {Paternostro}}, \ and\ \bibinfo {author} {\bibfnamefont {S.}~\bibnamefont {Gigan}},\ }\bibfield  {title} {\enquote {\bibinfo {title} {Programmable linear quantum networks with a multimode fibre},}\ }\href@noop {} {\bibfield  {journal} {\bibinfo  {journal} {Nature Photonics}\ }\textbf {\bibinfo {volume} {14}},\ \bibinfo {pages} {139--142} (\bibinfo {year} {2020})}\BibitemShut {NoStop}%
\bibitem [{\citenamefont {Goel}\ \emph {et~al.}(2024)\citenamefont {Goel}, \citenamefont {Leedumrongwatthanakun}, \citenamefont {Valencia}, \citenamefont {McCutcheon}, \citenamefont {Tavakoli}, \citenamefont {Conti}, \citenamefont {Pinkse},\ and\ \citenamefont {Malik}}]{goel2024inversedesign}%
  \BibitemOpen
  \bibfield  {author} {\bibinfo {author} {\bibfnamefont {S.}~\bibnamefont {Goel}}, \bibinfo {author} {\bibfnamefont {S.}~\bibnamefont {Leedumrongwatthanakun}}, \bibinfo {author} {\bibfnamefont {N.~H.}\ \bibnamefont {Valencia}}, \bibinfo {author} {\bibfnamefont {W.}~\bibnamefont {McCutcheon}}, \bibinfo {author} {\bibfnamefont {A.}~\bibnamefont {Tavakoli}}, \bibinfo {author} {\bibfnamefont {C.}~\bibnamefont {Conti}}, \bibinfo {author} {\bibfnamefont {P.~W.}\ \bibnamefont {Pinkse}}, \ and\ \bibinfo {author} {\bibfnamefont {M.}~\bibnamefont {Malik}},\ }\bibfield  {title} {\enquote {\bibinfo {title} {Inverse design of high-dimensional quantum optical circuits in a complex medium},}\ }\href@noop {} {\bibfield  {journal} {\bibinfo  {journal} {Nature Physics}\ ,\ \bibinfo {pages} {1--8}} (\bibinfo {year} {2024})}\BibitemShut {NoStop}%
\bibitem [{\citenamefont {Valencia}\ \emph {et~al.}(2024)\citenamefont {Valencia}, \citenamefont {Goel}, \citenamefont {Ma}, \citenamefont {Leedumrongwatthanakun}, \citenamefont {Graffitti}, \citenamefont {Fedrizzi}, \citenamefont {McCutcheon},\ and\ \citenamefont {Malik}}]{valencia2024multiplexed}%
  \BibitemOpen
  \bibfield  {author} {\bibinfo {author} {\bibfnamefont {N.~H.}\ \bibnamefont {Valencia}}, \bibinfo {author} {\bibfnamefont {S.}~\bibnamefont {Goel}}, \bibinfo {author} {\bibfnamefont {A.}~\bibnamefont {Ma}}, \bibinfo {author} {\bibfnamefont {S.}~\bibnamefont {Leedumrongwatthanakun}}, \bibinfo {author} {\bibfnamefont {F.}~\bibnamefont {Graffitti}}, \bibinfo {author} {\bibfnamefont {A.}~\bibnamefont {Fedrizzi}}, \bibinfo {author} {\bibfnamefont {W.}~\bibnamefont {McCutcheon}}, \ and\ \bibinfo {author} {\bibfnamefont {M.}~\bibnamefont {Malik}},\ }\bibfield  {title} {\enquote {\bibinfo {title} {A multiplexed programmable quantum photonic network},}\ }in\ \href@noop {} {\emph {\bibinfo {booktitle} {Quantum 2.0}}}\ (\bibinfo {organization} {Optica Publishing Group},\ \bibinfo {year} {2024})\ pp.\ \bibinfo {pages} {QTh2B--6}\BibitemShut {NoStop}%
\bibitem [{\citenamefont {Lahini}\ \emph {et~al.}(2010)\citenamefont {Lahini}, \citenamefont {Bromberg}, \citenamefont {Christodoulides},\ and\ \citenamefont {Silberberg}}]{lahini2010quantum}%
  \BibitemOpen
  \bibfield  {author} {\bibinfo {author} {\bibfnamefont {Y.}~\bibnamefont {Lahini}}, \bibinfo {author} {\bibfnamefont {Y.}~\bibnamefont {Bromberg}}, \bibinfo {author} {\bibfnamefont {D.~N.}\ \bibnamefont {Christodoulides}}, \ and\ \bibinfo {author} {\bibfnamefont {Y.}~\bibnamefont {Silberberg}},\ }\bibfield  {title} {\enquote {\bibinfo {title} {Quantum correlations in two-particle anderson localization},}\ }\href@noop {} {\bibfield  {journal} {\bibinfo  {journal} {Physical review letters}\ }\textbf {\bibinfo {volume} {105}},\ \bibinfo {pages} {163905} (\bibinfo {year} {2010})}\BibitemShut {NoStop}%
\bibitem [{\citenamefont {Black}\ \emph {et~al.}(2019)\citenamefont {Black}, \citenamefont {Giese}, \citenamefont {Braverman}, \citenamefont {Zollo}, \citenamefont {Barnett},\ and\ \citenamefont {Boyd}}]{black2019quantum}%
  \BibitemOpen
  \bibfield  {author} {\bibinfo {author} {\bibfnamefont {A.~N.}\ \bibnamefont {Black}}, \bibinfo {author} {\bibfnamefont {E.}~\bibnamefont {Giese}}, \bibinfo {author} {\bibfnamefont {B.}~\bibnamefont {Braverman}}, \bibinfo {author} {\bibfnamefont {N.}~\bibnamefont {Zollo}}, \bibinfo {author} {\bibfnamefont {S.~M.}\ \bibnamefont {Barnett}}, \ and\ \bibinfo {author} {\bibfnamefont {R.~W.}\ \bibnamefont {Boyd}},\ }\bibfield  {title} {\enquote {\bibinfo {title} {Quantum nonlocal aberration cancellation},}\ }\href@noop {} {\bibfield  {journal} {\bibinfo  {journal} {Physical Review Letters}\ }\textbf {\bibinfo {volume} {123}},\ \bibinfo {pages} {143603} (\bibinfo {year} {2019})}\BibitemShut {NoStop}%
\bibitem [{\citenamefont {Safadi}\ \emph {et~al.}(2023)\citenamefont {Safadi}, \citenamefont {Lib}, \citenamefont {Lin}, \citenamefont {Hsu}, \citenamefont {Goetschy},\ and\ \citenamefont {Bromberg}}]{safadi2023coherent}%
  \BibitemOpen
  \bibfield  {author} {\bibinfo {author} {\bibfnamefont {M.}~\bibnamefont {Safadi}}, \bibinfo {author} {\bibfnamefont {O.}~\bibnamefont {Lib}}, \bibinfo {author} {\bibfnamefont {H.-C.}\ \bibnamefont {Lin}}, \bibinfo {author} {\bibfnamefont {C.~W.}\ \bibnamefont {Hsu}}, \bibinfo {author} {\bibfnamefont {A.}~\bibnamefont {Goetschy}}, \ and\ \bibinfo {author} {\bibfnamefont {Y.}~\bibnamefont {Bromberg}},\ }\bibfield  {title} {\enquote {\bibinfo {title} {Coherent backscattering of entangled photon pairs},}\ }\href@noop {} {\bibfield  {journal} {\bibinfo  {journal} {Nature Physics}\ }\textbf {\bibinfo {volume} {19}},\ \bibinfo {pages} {562--568} (\bibinfo {year} {2023})}\BibitemShut {NoStop}%
\bibitem [{\citenamefont {Bajar}\ \emph {et~al.}(2025)\citenamefont {Bajar}, \citenamefont {Chatterjee}, \citenamefont {Bhat},\ and\ \citenamefont {Mujumdar}}]{bajar2025partial}%
  \BibitemOpen
  \bibfield  {author} {\bibinfo {author} {\bibfnamefont {K.}~\bibnamefont {Bajar}}, \bibinfo {author} {\bibfnamefont {R.}~\bibnamefont {Chatterjee}}, \bibinfo {author} {\bibfnamefont {V.~S.}\ \bibnamefont {Bhat}}, \ and\ \bibinfo {author} {\bibfnamefont {S.}~\bibnamefont {Mujumdar}},\ }\bibfield  {title} {\enquote {\bibinfo {title} {Partial immunity of two-photon correlation against wavefront distortion for spatially entangled photons},}\ }\href@noop {} {\bibfield  {journal} {\bibinfo  {journal} {APL Quantum}\ }\textbf {\bibinfo {volume} {2}} (\bibinfo {year} {2025})}\BibitemShut {NoStop}%
\bibitem [{\citenamefont {Monin}, \citenamefont {Alterman},\ and\ \citenamefont {Levin}(2025)}]{monin2025rapid}%
  \BibitemOpen
  \bibfield  {author} {\bibinfo {author} {\bibfnamefont {S.}~\bibnamefont {Monin}}, \bibinfo {author} {\bibfnamefont {M.}~\bibnamefont {Alterman}}, \ and\ \bibinfo {author} {\bibfnamefont {A.}~\bibnamefont {Levin}},\ }\bibfield  {title} {\enquote {\bibinfo {title} {Rapid wavefront shaping using an optical gradient acquisition},}\ }\href@noop {} {\bibfield  {journal} {\bibinfo  {journal} {arXiv preprint arXiv:2501.13711}\ } (\bibinfo {year} {2025})}\BibitemShut {NoStop}%
\bibitem [{\citenamefont {del Hougne}\ and\ \citenamefont {Lerosey}(2018)}]{del2018leveraging}%
  \BibitemOpen
  \bibfield  {author} {\bibinfo {author} {\bibfnamefont {P.}~\bibnamefont {del Hougne}}\ and\ \bibinfo {author} {\bibfnamefont {G.}~\bibnamefont {Lerosey}},\ }\bibfield  {title} {\enquote {\bibinfo {title} {Leveraging chaos for wave-based analog computation: Demonstration with indoor wireless communication signals},}\ }\href@noop {} {\bibfield  {journal} {\bibinfo  {journal} {Physical Review X}\ }\textbf {\bibinfo {volume} {8}},\ \bibinfo {pages} {041037} (\bibinfo {year} {2018})}\BibitemShut {NoStop}%
\bibitem [{\citenamefont {Xia}\ \emph {et~al.}(2024)\citenamefont {Xia}, \citenamefont {Kim}, \citenamefont {Eliezer}, \citenamefont {Han}, \citenamefont {Shaughnessy}, \citenamefont {Gigan},\ and\ \citenamefont {Cao}}]{xia2024nonlinear}%
  \BibitemOpen
  \bibfield  {author} {\bibinfo {author} {\bibfnamefont {F.}~\bibnamefont {Xia}}, \bibinfo {author} {\bibfnamefont {K.}~\bibnamefont {Kim}}, \bibinfo {author} {\bibfnamefont {Y.}~\bibnamefont {Eliezer}}, \bibinfo {author} {\bibfnamefont {S.}~\bibnamefont {Han}}, \bibinfo {author} {\bibfnamefont {L.}~\bibnamefont {Shaughnessy}}, \bibinfo {author} {\bibfnamefont {S.}~\bibnamefont {Gigan}}, \ and\ \bibinfo {author} {\bibfnamefont {H.}~\bibnamefont {Cao}},\ }\bibfield  {title} {\enquote {\bibinfo {title} {Nonlinear optical encoding enabled by recurrent linear scattering},}\ }\href@noop {} {\bibfield  {journal} {\bibinfo  {journal} {Nature Photonics}\ }\textbf {\bibinfo {volume} {18}},\ \bibinfo {pages} {1067--1075} (\bibinfo {year} {2024})}\BibitemShut {NoStop}%
\bibitem [{\citenamefont {Yildirim}\ \emph {et~al.}(2024)\citenamefont {Yildirim}, \citenamefont {Dinc}, \citenamefont {Oguz}, \citenamefont {Psaltis},\ and\ \citenamefont {Moser}}]{yildirim2024nonlinear}%
  \BibitemOpen
  \bibfield  {author} {\bibinfo {author} {\bibfnamefont {M.}~\bibnamefont {Yildirim}}, \bibinfo {author} {\bibfnamefont {N.~U.}\ \bibnamefont {Dinc}}, \bibinfo {author} {\bibfnamefont {I.}~\bibnamefont {Oguz}}, \bibinfo {author} {\bibfnamefont {D.}~\bibnamefont {Psaltis}}, \ and\ \bibinfo {author} {\bibfnamefont {C.}~\bibnamefont {Moser}},\ }\bibfield  {title} {\enquote {\bibinfo {title} {Nonlinear processing with linear optics},}\ }\href@noop {} {\bibfield  {journal} {\bibinfo  {journal} {Nature Photonics}\ }\textbf {\bibinfo {volume} {18}},\ \bibinfo {pages} {1076--1082} (\bibinfo {year} {2024})}\BibitemShut {NoStop}%
\end{thebibliography}%


\begin{thebibliography}{14}%
\makeatletter
\providecommand \@ifxundefined [1]{%
 \@ifx{#1\undefined}
}%
\providecommand \@ifnum [1]{%
 \ifnum #1\expandafter \@firstoftwo
 \else \expandafter \@secondoftwo
 \fi
}%
\providecommand \@ifx [1]{%
 \ifx #1\expandafter \@firstoftwo
 \else \expandafter \@secondoftwo
 \fi
}%
\providecommand \natexlab [1]{#1}%
\providecommand \enquote  [1]{``#1''}%
\providecommand \bibnamefont  [1]{#1}%
\providecommand \bibfnamefont [1]{#1}%
\providecommand \citenamefont [1]{#1}%
\providecommand \href@noop [0]{\@secondoftwo}%
\providecommand \href [0]{\begingroup \@sanitize@url \@href}%
\providecommand \@href[1]{\@@startlink{#1}\@@href}%
\providecommand \@@href[1]{\endgroup#1\@@endlink}%
\providecommand \@sanitize@url [0]{\catcode `\\12\catcode `\$12\catcode `\&12\catcode `\#12\catcode `\^12\catcode `\_12\catcode `\%12\relax}%
\providecommand \@@startlink[1]{}%
\providecommand \@@endlink[0]{}%
\providecommand \url  [0]{\begingroup\@sanitize@url \@url }%
\providecommand \@url [1]{\endgroup\@href {#1}{\urlprefix }}%
\providecommand \urlprefix  [0]{URL }%
\providecommand \Eprint [0]{\href }%
\providecommand \doibase [0]{http://dx.doi.org/}%
\providecommand \selectlanguage [0]{\@gobble}%
\providecommand \bibinfo  [0]{\@secondoftwo}%
\providecommand \bibfield  [0]{\@secondoftwo}%
\providecommand \translation [1]{[#1]}%
\providecommand \BibitemOpen [0]{}%
\providecommand \bibitemStop [0]{}%
\providecommand \bibitemNoStop [0]{.\EOS\space}%
\providecommand \EOS [0]{\spacefactor3000\relax}%
\providecommand \BibitemShut  [1]{\csname bibitem#1\endcsname}%
\let\auto@bib@innerbib\@empty
\bibitem [{\citenamefont {Howell}\ \emph {et~al.}(2004)\citenamefont {Howell}, \citenamefont {Bennink}, \citenamefont {Bentley},\ and\ \citenamefont {Boyd}}]{howell2004realization}%
  \BibitemOpen
  \bibfield  {author} {\bibinfo {author} {\bibfnamefont {J.~C.}\ \bibnamefont {Howell}}, \bibinfo {author} {\bibfnamefont {R.~S.}\ \bibnamefont {Bennink}}, \bibinfo {author} {\bibfnamefont {S.~J.}\ \bibnamefont {Bentley}}, \ and\ \bibinfo {author} {\bibfnamefont {R.~W.}\ \bibnamefont {Boyd}},\ }\bibfield  {title} {\enquote {\bibinfo {title} {Realization of the einstein-podolsky-rosen paradox using momentum- and position-entangled photons from spontaneous parametric down conversion},}\ }\href@noop {} {\bibfield  {journal} {\bibinfo  {journal} {Physical review letters}\ }\textbf {\bibinfo {volume} {92}},\ \bibinfo {pages} {210403} (\bibinfo {year} {2004})}\BibitemShut {NoStop}%
\bibitem [{\citenamefont {Walborn}\ \emph {et~al.}(2010)\citenamefont {Walborn}, \citenamefont {Monken}, \citenamefont {P{\'a}dua},\ and\ \citenamefont {Ribeiro}}]{walborn2010spatial}%
  \BibitemOpen
  \bibfield  {author} {\bibinfo {author} {\bibfnamefont {S.~P.}\ \bibnamefont {Walborn}}, \bibinfo {author} {\bibfnamefont {C.}~\bibnamefont {Monken}}, \bibinfo {author} {\bibfnamefont {S.}~\bibnamefont {P{\'a}dua}}, \ and\ \bibinfo {author} {\bibfnamefont {P.~S.}\ \bibnamefont {Ribeiro}},\ }\bibfield  {title} {\enquote {\bibinfo {title} {Spatial correlations in parametric down-conversion},}\ }\href@noop {} {\bibfield  {journal} {\bibinfo  {journal} {Physics Reports}\ }\textbf {\bibinfo {volume} {495}},\ \bibinfo {pages} {87--139} (\bibinfo {year} {2010})}\BibitemShut {NoStop}%
\bibitem [{\citenamefont {Gerry}\ and\ \citenamefont {Knight}(2023)}]{gerry2023introductory}%
  \BibitemOpen
  \bibfield  {author} {\bibinfo {author} {\bibfnamefont {C.~C.}\ \bibnamefont {Gerry}}\ and\ \bibinfo {author} {\bibfnamefont {P.~L.}\ \bibnamefont {Knight}},\ }\href@noop {} {\emph {\bibinfo {title} {Introductory quantum optics}}}\ (\bibinfo  {publisher} {Cambridge university press},\ \bibinfo {year} {2023})\BibitemShut {NoStop}%
\bibitem [{\citenamefont {Beenakker}(1997)}]{beenakker1997random}%
  \BibitemOpen
  \bibfield  {author} {\bibinfo {author} {\bibfnamefont {C.~W.}\ \bibnamefont {Beenakker}},\ }\bibfield  {title} {\enquote {\bibinfo {title} {Random-matrix theory of quantum transport},}\ }\href@noop {} {\bibfield  {journal} {\bibinfo  {journal} {Reviews of modern physics}\ }\textbf {\bibinfo {volume} {69}},\ \bibinfo {pages} {731} (\bibinfo {year} {1997})}\BibitemShut {NoStop}%
\bibitem [{\citenamefont {Mastrodonato}\ and\ \citenamefont {Tumulka}(2007)}]{mastrodonato2007elementary}%
  \BibitemOpen
  \bibfield  {author} {\bibinfo {author} {\bibfnamefont {C.}~\bibnamefont {Mastrodonato}}\ and\ \bibinfo {author} {\bibfnamefont {R.}~\bibnamefont {Tumulka}},\ }\bibfield  {title} {\enquote {\bibinfo {title} {Elementary proof for asymptotics of large haar-distributed unitary matrices},}\ }\href@noop {} {\bibfield  {journal} {\bibinfo  {journal} {Letters in Mathematical Physics}\ }\textbf {\bibinfo {volume} {82}},\ \bibinfo {pages} {51--59} (\bibinfo {year} {2007})}\BibitemShut {NoStop}%
\bibitem [{\citenamefont {Goodman}(2007)}]{goodman2007speckle}%
  \BibitemOpen
  \bibfield  {author} {\bibinfo {author} {\bibfnamefont {J.~W.}\ \bibnamefont {Goodman}},\ }\href@noop {} {\emph {\bibinfo {title} {Speckle phenomena in optics: theory and applications}}}\ (\bibinfo  {publisher} {Roberts and Company Publishers},\ \bibinfo {year} {2007})\BibitemShut {NoStop}%
\bibitem [{\citenamefont {Arora}\ \emph {et~al.}(2005)\citenamefont {Arora}, \citenamefont {Berger}, \citenamefont {Elad}, \citenamefont {Kindler},\ and\ \citenamefont {Safra}}]{arora2005non}%
  \BibitemOpen
  \bibfield  {author} {\bibinfo {author} {\bibfnamefont {S.}~\bibnamefont {Arora}}, \bibinfo {author} {\bibfnamefont {E.}~\bibnamefont {Berger}}, \bibinfo {author} {\bibfnamefont {H.}~\bibnamefont {Elad}}, \bibinfo {author} {\bibfnamefont {G.}~\bibnamefont {Kindler}}, \ and\ \bibinfo {author} {\bibfnamefont {M.}~\bibnamefont {Safra}},\ }\bibfield  {title} {\enquote {\bibinfo {title} {On non-approximability for quadratic programs},}\ }in\ \href@noop {} {\emph {\bibinfo {booktitle} {46th Annual IEEE Symposium on Foundations of Computer Science (FOCS'05)}}}\ (\bibinfo {organization} {IEEE},\ \bibinfo {year} {2005})\ pp.\ \bibinfo {pages} {206--215}\BibitemShut {NoStop}%
\bibitem [{\citenamefont {Pardalos}\ and\ \citenamefont {Vavasis}(1991)}]{pardalos1991quadratic}%
  \BibitemOpen
  \bibfield  {author} {\bibinfo {author} {\bibfnamefont {P.~M.}\ \bibnamefont {Pardalos}}\ and\ \bibinfo {author} {\bibfnamefont {S.~A.}\ \bibnamefont {Vavasis}},\ }\bibfield  {title} {\enquote {\bibinfo {title} {Quadratic programming with one negative eigenvalue is np-hard},}\ }\href@noop {} {\bibfield  {journal} {\bibinfo  {journal} {Journal of Global optimization}\ }\textbf {\bibinfo {volume} {1}},\ \bibinfo {pages} {15--22} (\bibinfo {year} {1991})}\BibitemShut {NoStop}%
\bibitem [{\citenamefont {Lucas}(2014)}]{lucas2014ising}%
  \BibitemOpen
  \bibfield  {author} {\bibinfo {author} {\bibfnamefont {A.}~\bibnamefont {Lucas}},\ }\bibfield  {title} {\enquote {\bibinfo {title} {Ising formulations of many np problems},}\ }\href@noop {} {\bibfield  {journal} {\bibinfo  {journal} {Frontiers in physics}\ }\textbf {\bibinfo {volume} {2}},\ \bibinfo {pages} {5} (\bibinfo {year} {2014})}\BibitemShut {NoStop}%
\bibitem [{\citenamefont {Pierangeli}, \citenamefont {Marcucci},\ and\ \citenamefont {Conti}(2019)}]{pierangeli2019large}%
  \BibitemOpen
  \bibfield  {author} {\bibinfo {author} {\bibfnamefont {D.}~\bibnamefont {Pierangeli}}, \bibinfo {author} {\bibfnamefont {G.}~\bibnamefont {Marcucci}}, \ and\ \bibinfo {author} {\bibfnamefont {C.}~\bibnamefont {Conti}},\ }\bibfield  {title} {\enquote {\bibinfo {title} {Large-scale photonic ising machine by spatial light modulation},}\ }\href@noop {} {\bibfield  {journal} {\bibinfo  {journal} {Physical review letters}\ }\textbf {\bibinfo {volume} {122}},\ \bibinfo {pages} {213902} (\bibinfo {year} {2019})}\BibitemShut {NoStop}%
\bibitem [{\citenamefont {Pierangeli}\ \emph {et~al.}(2021)\citenamefont {Pierangeli}, \citenamefont {Rafayelyan}, \citenamefont {Conti},\ and\ \citenamefont {Gigan}}]{pierangeli2021scalable}%
  \BibitemOpen
  \bibfield  {author} {\bibinfo {author} {\bibfnamefont {D.}~\bibnamefont {Pierangeli}}, \bibinfo {author} {\bibfnamefont {M.}~\bibnamefont {Rafayelyan}}, \bibinfo {author} {\bibfnamefont {C.}~\bibnamefont {Conti}}, \ and\ \bibinfo {author} {\bibfnamefont {S.}~\bibnamefont {Gigan}},\ }\bibfield  {title} {\enquote {\bibinfo {title} {Scalable spin-glass optical simulator},}\ }\href@noop {} {\bibfield  {journal} {\bibinfo  {journal} {Physical Review Applied}\ }\textbf {\bibinfo {volume} {15}},\ \bibinfo {pages} {034087} (\bibinfo {year} {2021})}\BibitemShut {NoStop}%
\bibitem [{\citenamefont {Courme}\ \emph {et~al.}(2025)\citenamefont {Courme}, \citenamefont {Verni{\`e}re}, \citenamefont {Joly}, \citenamefont {Faccio}, \citenamefont {Gigan},\ and\ \citenamefont {Defienne}}]{courme2025non}%
  \BibitemOpen
  \bibfield  {author} {\bibinfo {author} {\bibfnamefont {B.}~\bibnamefont {Courme}}, \bibinfo {author} {\bibfnamefont {C.}~\bibnamefont {Verni{\`e}re}}, \bibinfo {author} {\bibfnamefont {M.}~\bibnamefont {Joly}}, \bibinfo {author} {\bibfnamefont {D.}~\bibnamefont {Faccio}}, \bibinfo {author} {\bibfnamefont {S.}~\bibnamefont {Gigan}}, \ and\ \bibinfo {author} {\bibfnamefont {H.}~\bibnamefont {Defienne}},\ }\bibfield  {title} {\enquote {\bibinfo {title} {Non-classical optimization through complex media},}\ }\href@noop {} {\bibfield  {journal} {\bibinfo  {journal} {arXiv preprint arXiv:2503.24283}\ } (\bibinfo {year} {2025})}\BibitemShut {NoStop}%
\bibitem [{qwf()}]{qwfs2024}%
  \BibitemOpen
  \href@noop {} {}\bibinfo {howpublished} {\url{https://doi.org/10.5281/zenodo.15300939}},\ \bibinfo {note} {code and data for publication: "Fundamental bounds of Wavefront Shaping of Spatially Entangled Photons"}\BibitemShut {NoStop}%
\bibitem [{\citenamefont {Vellekoop}\ and\ \citenamefont {Mosk}(2008)}]{vellekoop2008universal}%
  \BibitemOpen
  \bibfield  {author} {\bibinfo {author} {\bibfnamefont {I.~M.}\ \bibnamefont {Vellekoop}}\ and\ \bibinfo {author} {\bibfnamefont {A.~P.}\ \bibnamefont {Mosk}},\ }\bibfield  {title} {\enquote {\bibinfo {title} {Universal optimal transmission of light through disordered materials},}\ }\href@noop {} {\bibfield  {journal} {\bibinfo  {journal} {Physical review letters}\ }\textbf {\bibinfo {volume} {101}},\ \bibinfo {pages} {120601} (\bibinfo {year} {2008})}\BibitemShut {NoStop}%
\end{thebibliography}%


\end{document}


\title{Fundamental Bounds of Wavefront Shaping of Spatially Entangled Photons - Supplementary Information} 

\author{Ronen Shekel}
\affiliation{Racah Institute of Physics,The Hebrew University of Jerusalem, Jerusalem, 91904, Israel}
\author{Sébastien M. Popoff}
\affiliation{Institut Langevin, ESPCI Paris, PSL University, CNRS, France}
\author{Yaron Bromberg}
\email[]{Yaron.Bromberg@mail.huji.ac.il}
\affiliation{Racah Institute of Physics,The Hebrew University of Jerusalem, Jerusalem, 91904, Israel}

\date{\today}

\pacs{}

\maketitle 

\makeatletter
\renewcommand \thesection{S\@arabic\c@section}
\renewcommand\thetable{S\@arabic\c@table}
\makeatother

\def\thefigure{S\arabic{figure}}
\setcounter{figure}{0}
\renewcommand{\theequation}{S\arabic{equation}}
\setcounter{equation}{0}

\section{Derivation of the two-photon probability $P_{\alpha\beta}$} \label{SI:klsyhko}
In this section, we derive Eq.~(1) from the main text, which gives the probability of a coincidence detection between photons in modes $\alpha$ and $\beta$ for the three configurations studied in this work. We also highlight its classical analog through the advanced wave picture.

We assume the state emitted by SPDC is a maximally entangled EPR-like state, represented in the position basis by $\left|\Psi_{\text{in}}\right\rangle =\frac{1}{\sqrt{2N}}\sum_{n}\hat{a}_{n}^{\dagger}\hat{a}_{n}^{\dagger}\left|\text{vac}\right\rangle$ ~\cite{howell2004realization}, where $\hat{a}_{n}^{\dagger}$ is the creation operator at mode $n$ and $|\text{vac}\rangle$ denotes the vacuum state. This state is obtained at the output facet of the nonlinear crystal in the so-called “thin crystal regime” ~\cite{walborn2010spatial}, and reflects the fact that the two photons emerge in the same spatial mode defined by the phase-matching conditions, and in a superposition of all transverse modes within the interaction volume. 

\begin{figure}[ht!]
    \centering
    \includegraphics[width=\linewidth]{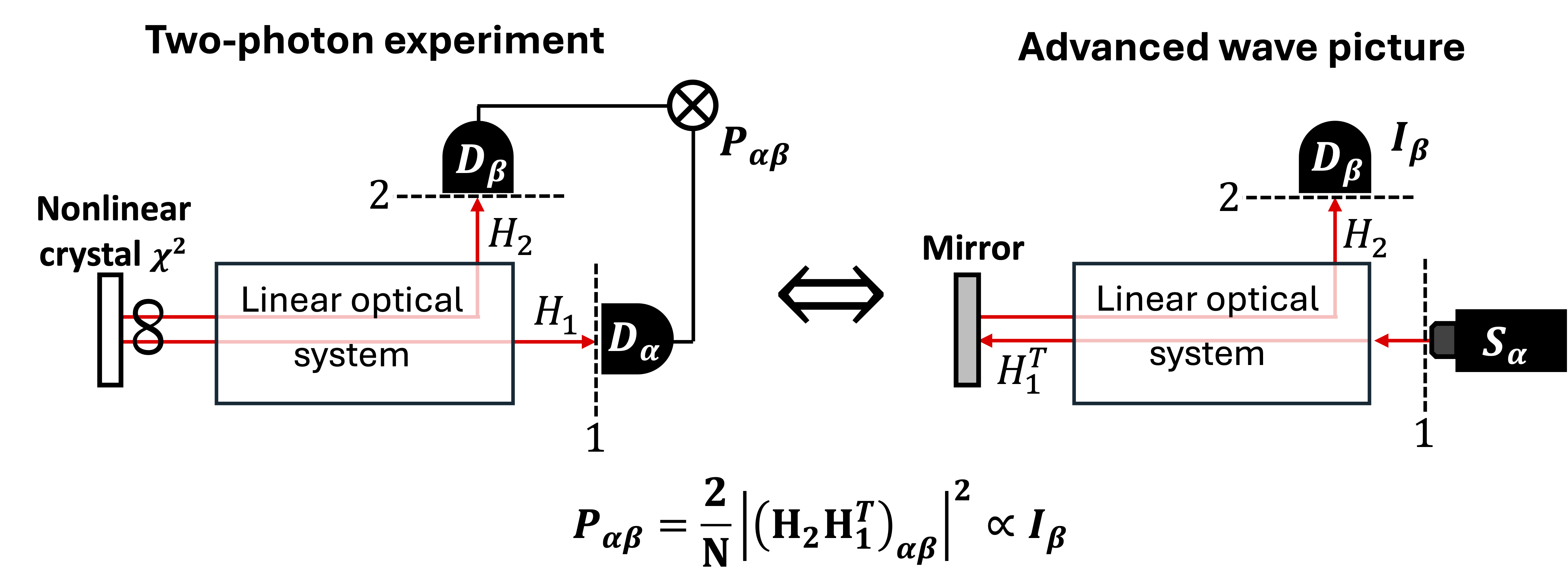} 
    \caption{\label{fig:AWP} The advanced wave interpretation of two-photon correlations between spatially entangled photons. For any reciprocal linear optical setup, where spatially entangled photons generated via SPDC propagate through linear optical systems $H_1, H_2$, respectively, the two-photon detection probability $P_{\alpha\beta}$ is equivalent to the average intensity $I_\beta$ of classical light in an advanced wave configuration. In the advanced wave picture, a classical laser originating from mode $\alpha$ at plane $1$, propagates backward through the optical setup $H_1$, reflects off a mirror in the crystal plane, and then propagates through the optical setup $H_2$ before finally being measured in mode $\beta$ at plane $2$. Since backpropagation is described by the transpose of the corresponding transmission matrix, the intensity at mode $\beta$ is given by $I_{\beta}\propto |(H_2H^{\mathrm{T}}_1)_{\alpha\beta}|^2$. Thus from Eq.~(\ref{eq:Pab_AWP}) we get $I_{\beta} \propto P_{\alpha\beta}$.}
\end{figure}

We denote by $H^{(1)}$ and $H^{(2)}$ the transmission matrices that describe the propagation from the crystal plane, to the plane of detectors $1$ and $2$, respectively, as depicted in Fig.~\ref{fig:AWP}. In the quantum mechanical description of the photon propagation, the matrix elements $h^{(i)}_{kn}$ denotes the probability amplitude for a photon to propagate from mode $n$ in the crystal plane to mode $k$ in the plane of detector $i$ ~\cite{gerry2023introductory}. The propagation of the creation operators is thus described by $\hat{a}_{n}^{\dagger}\rightarrow\sum_{k=1}^{N}h^{(1)}_{kn}\hat{a}_{k}^{(1)\dagger}+h^{(2)}_{kn}\hat{a}_{k}^{(2)\dagger}$, where $\hat{a}_{n}^{(i)\dagger}$ is the creation operator of mode $n$ in the plane of detector $i$. The state at the detector plane is thus described by:

\begin{equation}
\label{eq:psi_out}
    \left|\Psi_{\text{in}}\right\rangle \rightarrow\left|\Psi_{\text{out}}\right \rangle= \frac{1}{\sqrt{2N}}\sum_{n,k,l=1}^N\left(h^{(1)}_{kn}{\hat{a}_k^{(1)\dagger}}+h^{(2)}_{kn}{\hat{a}_k^{(2)\dagger}}\right) \left(h^{(1)}_{ln}{\hat{a}_l^{(1)\dagger}}+h^{(2)}_{ln}{\hat{a}_l^{(2)\dagger}}\right)\left|\text{vac}\right\rangle.
\end{equation}

The probability of measuring a coincidence event of detection of a photon in mode $\alpha$ by detector $1$ and a photon in mode $\beta$ by detector $2$ is given by:
\begin{equation}
\label{eq:Pab}
P_{\alpha\beta}=\left|\langle \text{vac}|\hat{a}^{(1)}_{\alpha}\hat{a}^{(2)}_{\beta}\left|\Psi_{\text{out}}\right \rangle\right|^2=\frac{2}{N}\left|\sum_{n}h^{(1)}_{\alpha n}h^{(2)}_{\beta n}\right|^2,
\end{equation}
where we used $\langle\text{vac}|\hat{a}^{(1)}_\alpha\hat{a}^{(2)}_\beta\hat{a}_k^{(i)\dagger}\hat{a}_l^{(j)\dagger}|\text{vac}\rangle=\delta_{i,1}\delta_{\alpha,k}\delta_{j,2}\delta_{\beta,l}+\delta_{i,2}\delta_{\alpha,l}\delta_{j,1}\delta_{\beta,k}$.

Since $h^{(i)}_{mn}$ describes the amplitude for a photon to propagate from mode $n$ in the crystal plane to mode $m$ of the plane of detector $i$, we can interpret Eq.~(\ref{eq:Pab}) as the coherent sum of all amplitudes corresponding to a pair of photons propagating from mode $n$ to mode $\alpha$ of detector $1$ and mode $\beta$ of detector $2$. Denoting the transpose of the matrix $H^{(1)}$ by $H^{(1)\mathsf{T}}$, we can write Eq.~(\ref{eq:Pab}) in a matrix form:
\begin{equation}
\label{eq:Pab_AWP}
P_{\alpha\beta}=\frac{2}{N}\left|\left(H^{(2)}H^{(1)\mathsf{T}}\right)_{\beta\alpha}\right|^2. 
\end{equation}

While Eq.~(\ref{eq:Pab_AWP}) was derived for the probability of a coincidence event between a pair of photons initially in an EPR state, we can also assign a classical interpretation to the right-hand side of the equation. It corresponds to the intensity measured in mode $\beta$ at plane $2$ when a classical field is launched from mode $\alpha$ at plane $1$ and propagated through a system described by the transmission matrix $H^{(2)}H^{(1)\mathsf{T}}$ (Fig. \ref{fig:AWP}). By optical reciprocity, $H^{\mathsf{T}}$ corresponds to propagation through the same optical system described by $H$, but in the reverse direction. Eq.~(\ref{eq:Pab_AWP}) thus illustrates the advanced wave picture: for an EPR input state $|\Psi_{\text{in}}\rangle$ emitted by the nonlinear crystal, the probability of a coincidence event between modes $\alpha$ and $\beta$ in planes $1$ and $2$ respectively, is proportional to the intensity measured when a classical field is emitted from mode $\alpha$ at plane $1$, propagates backward through the optical system toward the crystal, is reflected by the crystal, and then propagates forward through the system and detected in mode $\beta$ at plane $2$.

We can now derive the two-photon probabilities for the three configurations studied in this work. For the 1P-S configuration, shown in Fig.~(1a) and its advanced wave picture in Fig.~(2a), only one of the photons is modulated by the SLM. The transmission matrices describing the optical paths of the two photons are given by $H_1 = \frac{1}{\sqrt{2}}\mathcal{F}ST$ and $H_2 = \frac{1}{\sqrt{2}}\mathcal{F}T$, where $T$ is the transmission matrix of the scattering sample, $S$ is a diagonal matrix representing the action of the SLM in the pixel mode basis, $\mathcal{F}$ is the discrete Fourier transform matrix representing the propagation to the detectors' plane in the far-field, and the $\frac{1}{\sqrt{2}}$ factor results from the beam splitter placed in front of the detectors. The two-photon coincidence probability for this configuration is therefore given by

\begin{equation}
\label{eq:Pab_1PS}
P^{(1P-S)}_{\alpha\beta} = \frac{1}{2N} \left| \left( \mathcal{F}TT^{\mathsf{T}}S\mathcal{F} \right)_{\beta\alpha} \right|^2.
\end{equation}
Here we have used the fact that both $S$ and $\mathcal{F}$ are symmetric matrices and are equal to their transpose.

For the 2P-IS configuration, shown in Fig.~(1b) and its advanced wave picture in Fig.~(2b), both photons propagate through the same optical elements. The transmission matrix of each photon is given by $H=\frac{1}{\sqrt{2}}\mathcal{F}TS$, resulting in the following two-photon coincidence probability: 

\begin{equation}
\label{eq:Pab_2PIS}
P^{(2P-IS)}_{\alpha\beta} = \frac{1}{2N} \left| \left( \mathcal{F}TSST^{\mathsf{T}}\mathcal{F} \right)_{\beta\alpha} \right|^2.
\end{equation}

Finally, for the 2P-DS configuration, shown in Fig.~(1c) and its advanced wave picture in Fig.~(2c), the transmission matrix of each photon is given by $H=\mathcal{F}ST$ and the corresponding two-photon probability is therefore: 

\begin{equation}
\label{eq:Pab_2PDS}
P^{(2P-DS)}_{\alpha\beta} = \frac{1}{2N} \left| \left(\mathcal{F}STT^{\mathsf{T}}S\mathcal{F} \right)_{\beta\alpha} \right|^2.
\end{equation}

\section{Analytical derivation of the 1P-S and 2P-IS enhancements} \label{SI:enhancement}
In this section, we present analytical expressions for the optimal enhancement factors achieved in the 1P-S and 2P-IS configurations. The derivation assumes that the elements of $T$ are complex circular Gaussian random variables. However, to leading order in $1/N$, the moments of the matrix elements of unitary matrices match the moments of the Gaussian model ~\cite{beenakker1997random, mastrodonato2007elementary}, so the results apply similarly to both models. 

We start by explicitly writing the sum in Eq.~(\ref{eq:Pab}):

\begin{equation}
\label{eq:P_explicit}
P_{\alpha\beta}=\frac{2}{N}\left|\sum_n h^{(2)}_{\beta n} h^{(1)}_{\alpha n}\right|^2=\frac{2}{N}\sum_n\left|h^{(2)}_{\beta n}\right|^2\left|h^{(1)}_{\alpha n}\right|^2+\frac{2}{N}\sum_{n\neq n'}h^{(2)}_{\beta n} h^{(1)}_{\alpha n}{h^{(2)}_{\beta n'}}^*{h^{(1)}_{\alpha n'}}^*.
\end{equation}

To compute the predicted enhancement factors, we need to find the probability $P_{\alpha\beta}$ before and after the SLM optimization. To find the probability before the optimization, we set the transmission matrix of the SLM to identity $S=I$, so that for all configurations we get $H_1=H_2=\frac{1}{\sqrt{2}}\mathcal{F}T\equiv\frac{1}{\sqrt{2}}\tilde{T}$, and substitute its matrix elements $\tilde{t}_{nm}$ into Eq.~(\ref{eq:P_explicit}).  According to our definition of the enhancement, we first evaluate the pre-optimization probability by performing disorder and spatial averaging, denoted by $\langle\!\langle \cdot \rangle\!\rangle$. This averaged value serves as the reference against which we compare the optimized probability at modes $\alpha$ and $\beta$, thereby quantifying the enhancement relative to the surrounding modes. We thus obtain:

\begin{equation}
\label{eq:P0_1ps}
P_0=\frac{1}{2}\langle\!\langle|\tilde{t}_{\alpha m}|^2|\tilde{t}_{\beta m}|^2\rangle\!\rangle+\frac{1}{2N}\sum_{n\neq n'}\langle\!\langle \tilde{t}_{\beta n} \tilde{t}_{\alpha n}{\tilde{t}}^{*}_{\beta n'}\tilde{t}^{*}_{\alpha n'}\rangle\!\rangle.
\end{equation}

Since the elements of $T$ are complex circular Gaussian random variables, and since such statistics are preserved under linear transformations ~\cite{goodman2007speckle}, the elements of $\tilde{T}$ are also complex circular Gaussian random variables. Applying the Gaussian moment theorem ~\cite{goodman2007speckle} we obtain $\langle\!\langle \tilde{t}_{\beta n} \tilde{t}_{\alpha n}{\tilde{t}^{*}_{\beta n'}}{\tilde{t}^{*}_{\alpha n'}}\rangle\!\rangle=
\langle\!\langle \tilde{t}_{\alpha n} {\tilde{t}^{*}_{\alpha n'}}\rangle\!\rangle\langle\!\langle \tilde{t}_{\beta n}{\tilde{t}^{*}_{\beta n'}}\rangle\!\rangle+
\langle\!\langle \tilde{t}_{\alpha n} {\tilde{t}^{*}_{\beta n'}}\rangle\!\rangle\langle\!\langle \tilde{t}_{\beta n}{\tilde{t}^{*}_{\alpha n'}}\rangle\!\rangle=0$, where in the last step we used the assumption that the columns of $T$ are uncorrelated, and hence the Fourier transforms of the columns are also uncorrelated, namely $\langle\!\langle \tilde{t}_{\beta n}{\tilde{t}^{*}_{\alpha n'}}\rangle\!\rangle=\langle\!\langle t_{\beta n}{t^*}_{\alpha n'}\rangle\!\rangle=0$ for $n\neq n'$. 
We thus get
\begin{equation}
    \label{eq:P0_T}
P_0=\frac{1}{2}\langle\!\langle|\tilde{t}_{\alpha m}|^2|\tilde{t}_{\beta m}|^2\rangle\!\rangle\approx\frac{1}{2}\langle\!\langle|\tilde{t}_{\alpha m}|^2\rangle\!\rangle^2,
\end{equation}

where in the last step we assumed that, due to the spatial averaging, the dominant contributions come from modes with $\alpha \neq \beta$, and hence the two matrix elements are uncorrelated. 
We can also compute $P_0$ directly from the definition of $P_{\alpha\beta}$ [Eq.~\ref{eq:Pab_AWP}] for $H_1=H_2=\frac{1}{\sqrt{2}}\tilde{T}$:
\begin{equation}
\label{eq:P0_TT}
P_0=\frac{1}{2N}\langle\!\langle|(\tilde{T}\tilde{T}^\mathsf{T})_{\beta\alpha}|^2\rangle\!\rangle = \frac{1}{2N}\langle\!\langle|\tilde{\tilde{t}}_{\beta\alpha}|^2\rangle\!\rangle,
\end{equation}
where $\tilde{\tilde{t}}_{\beta\alpha}$ is the matrix element of $\tilde{T}\tilde{T}^\mathsf{T}$. 

From Eq.~(\ref{eq:P0_T}) and Eq.~(\ref{eq:P0_TT}) we get that  $\langle\!\langle|\tilde{\tilde{t}}_{mn}|^2\rangle\!\rangle=N\langle\!\langle|\tilde{t}_{ mn}|^2\rangle\!\rangle^2$, which is consistent with our assumption that the matrix elements of each transmission matrix have the same average intensity, so that that if the average total transmission through the matrix $\tilde{T}$ is $\tau=N\langle\!\langle|\tilde{t}_{mn}|^2\rangle\!\rangle$, then the average matrix element of $\tilde{T}\tilde{T}^{\mathsf{T}}$, which describes a double pass through $\tilde{T}$, is given by  $\langle\!\langle|\tilde{\tilde{t}}_{ mn}|^2\rangle\!\rangle=\tau^2/N$.

For the optimal probability that can be achieved using the SLM, we note that a global maximum of the probability can be obtained if the SLM is used to set the same phase for all terms in the sum in Eq.~(\ref{eq:P_explicit}). Whether this is possible, in principle, depends on the location of the SLM. Below, we show that this is indeed possible in the 1P-S and 2P-IS configurations.

\subsection{1P-S configuration}
\label{sec:1PS}
For the 1P-S configuration, $H_1=\frac{1}{\sqrt{2}}\mathcal{F}ST$ and $H_2=\frac{1}{\sqrt{2}}\tilde{T}$. To show that we can use the SLM to make all the terms in the  sum in Eq.~(\ref{eq:P_explicit}) in phase, we write the sum explicitly

\begin{align}
    \sum_n h^{(2)}_{\beta n} h^{(1)}_{\alpha n} 
    &= \frac{1}{2}\sum_n {\tilde{t}}_{\beta n} \sum_m f_{\alpha m} e^{i\Phi_m} t_{mn} \notag 
    = \frac{1}{2}\sum_m f_{\alpha m} e^{i\Phi_m} \sum_n {\tilde{t}}_{\beta n} t_{mn} \notag \\
    &= \frac{1}{2}\sum_m f_{\alpha m} e^{i\Phi_m}t'_{\beta m},\label{eq:sum_1PS}
\end{align}

where we keep the notation in which lowercase elements $f_{mn}$, $t_{mn}$ and $t'_{mn}$ correspond to the uppercase matrices $\mathcal{F}$, $T$ and $T'\equiv \tilde{T}T^{\mathsf{T}}$, respectively. The term $e^{i\Phi_m}$ denotes the $m^{\text{th}}$ matrix element of the SLM diagonal matrix $S$, representing the phase applied by pixel $m$ of the SLM.  By setting the SLM phases to be $\Phi_m=-\text{arg}(f_{\alpha m})-\text{arg}(t'_{\beta m})$, where $\text{arg()}$ denotes the phase of the complex matrix element, all terms in the sum add in phase. As a result, the right hand side of Eq.~(\ref{eq:sum_1PS}) becomes  $\frac{1}{2}\sum_m\frac{1}{\sqrt{N}}|t'_{\beta m}|$, where we used the fact that the modulus of the Fourier matrix element satisfies $|f_{\alpha n}|=\frac{1}{\sqrt{N}}$. 
The disorder averaged optimal probability is thus given by:
\begin{equation}
\label{eq:Popt_1ps}
P_{\text{opt}}=\frac{1}{2N}\left\langle\left(\sum_m\frac{|t'_{\beta m}|}{\sqrt{N}}\right)^2\right\rangle=\frac{1}{2N}\left[ \left\langle|t'_{\beta m}|^2\right\rangle+(N-1)\left\langle|t'_{\beta m}|\right\rangle^2\right].
\end{equation}

Here, $\langle \cdot \rangle$ denotes disorder averaging. We have used the fact that the disorder average of all matrix elements is identical, and that since the columns of $T$ are uncorrelated, the columns of $\tilde{T}$ are also uncorrelated. As a result, $\langle |t'_{\alpha m}|\,|t'_{\alpha n}| \rangle = \langle |t'_{\alpha m}| \rangle \langle |t'_{\alpha n}| \rangle$ for $m \neq n$. Using Eq.(\ref{eq:Popt_1ps}) and Eq.(\ref{eq:P0_TT}), along with the fact that $\mathcal{F}$ is a unitary transformation, so that $\langle|\tilde{\tilde{t}}_{mn}|^2\rangle = \langle|t’_{mn}|^2\rangle$, we find that the enhancement for the 1P-S configuration is given by:

\begin{equation}
\label{eq:eta_1ps}
    \eta_{1P-S}=\frac{P_{\text{opt}}}{P_0}=\frac{ \left\langle|t'_{mn}|^2\right\rangle+(N-1)\left\langle|t'_{mn}|\right\rangle^2}{\langle\!\langle|t'_{mn}|^2\rangle\!\rangle}=1+\frac{\pi}{4}(N-1)\mathrel{\underset{N \gg 1}{\longrightarrow}}\frac{\pi}{4}N. 
\end{equation}

Here we assumed that for $N\gg 1$ the elements of $T’$ are complex circular Gaussian random variables, leading to $\langle |t’_{mn}| \rangle^2 / \langle |t’_{mn}|^2 \rangle = \pi/4$ ~\cite{goodman2007speckle}.

\subsection{2P-IS configuration}
For the 2P-IS configuration, $H_1=H_2=\frac{1}{\sqrt{2}}\mathcal{F}TS=\frac{1}{\sqrt{2}}\tilde{T}
S$. The  sum in Eq.~(\ref{eq:P_explicit}) is then

\begin{align}
    \sum_n h^{(2)}_{\beta n} h^{(1)}_{\alpha n} 
    &= \frac{1}{2}\sum_n \tilde{t}_{\beta n}\tilde{t}_{\alpha n} e^{i2\Phi_n}.\label{eq:sum_2PS}
\end{align}

By setting the SLM phases to be $\Phi_n=-\frac{1}{2}[\text{arg}(\tilde{t}_{\alpha n})+\text{arg}(\tilde{t}_{\beta n})]$, all terms in the sum add in phase and we get: 

\begin{equation}
\label{eq:Popt_2ps}
P_{\text{opt}}=\frac{1}{2N}\left\langle\left|\sum_n|\tilde{t}_{\alpha n}||\tilde{t}_{\beta n}|\right|^2\right\rangle= \frac{1}{2}\left[ \left\langle|\tilde{t}_{\alpha n}|^2|\tilde{t}_{\beta n}|^2\right\rangle+(N-1)\left\langle|\tilde{t}_{\alpha n}| |\tilde{t}_{\beta n}|\right\rangle^2\right].
\end{equation}

To get the enhancement for the 2P-IS configuration, we divide Eq.~(\ref{eq:Popt_2ps}) by Eq.~(\ref{eq:P0_T}). Assuming the coincidence measurement is between two different modes ($\alpha\neq \beta$), we get: 

\begin{equation}
\label{eq:eta_2ps}
    \eta_{2P-IS}=\frac{P_{\text{opt}}}{P_0}=\frac{ \left\langle|\tilde{t}_{mn}|^2\right\rangle^2+(N-1)\left\langle|\tilde{t}_{mn}|\right\rangle^4}{{\langle|\tilde{t}_{mn}|^2\rangle}^2}=1+\left(\frac{\pi}{4}\right)^2(N-1)\mathrel{\underset{N \gg 1}{\longrightarrow}} \left(\frac{\pi}{4}\right)^2 N,
\end{equation}

where we used the fact that the elements of $\tilde{T}$ are complex circular Gaussian random variables, implying $\langle |\tilde{t}_{mn}| \rangle^2 / \langle |\tilde{t}_{mn}|^2 \rangle = \pi/4$, as in the previous case.

\subsubsection*{Digital Optical Phase Conjugation (OPC)}
In the special case in which the two detectors are located in the same transverse position, such that they detect the same spatial mode ($\alpha=\beta$), we get that the enhancement is: 

\begin{equation}
\label{eq:eta_2ps_OPC}
    \eta^{(OPC)}_{2P-IS}=\frac{ \left\langle|\tilde{t}_{mn}|^4\right\rangle+(N-1)\left\langle|\tilde{t}_{mn}|^2\right\rangle^2}{{\langle|\tilde{t}_{mn}|^2\rangle}^2}=2+(N-1) \mathrel{\underset{N \gg 1}{\longrightarrow}} N,
\end{equation}
where we have used the fact that for complex circular Gaussian variables $\left\langle|\tilde{t}_{mn}|^4\right\rangle=2\left\langle|\tilde{t}_{mn}|^2\right\rangle^2$. 

We refer to this configuration as \textit{digital optical phase conjugation}, since in this configuration the optimal probability $P_{\text{opt}}$ is according to Eq.~(\ref{eq:Popt_2ps}) $P_{\text{opt}}=\frac{1}{2N}\langle|\sum_n|\tilde{t}_{\alpha n}|^2|^2\rangle=\tau^2/2N$ where $\tau$ denotes the total transmission through the scattering sample. For a lossless sample, the transmission matrix is unitary so that $\tau=1$. In this case, the optimal probability equals the probability of a coincidence event in the absence of the scattering sample and with perfect imaging of the two-photon source onto the detector plane. In addition, in the unitary case $\sum_{n=1}\left|t_{mn}\right|^{2}=1$ exactly, so from Eq. (\ref{eq:Popt_2ps}) we get $P_{\text{opt}}=1/2N$ and the enhancement is $\eta^{(OPC)}_{2P-IS}=N$.

\section{Computational complexity of the 2P-DS configuration}\label{SI:NP-hard}
The optimization task in the 2P-DS configuration involves maximizing the coincidence probability $P^{(2P-DS)}_{\alpha\beta}$ given in Eq.~(1) of the main text:

\begin{equation}
    P^{(2P-DS)}_{\alpha\beta} = \frac{1}{2N} \left| \left(\mathcal{F}STT^{\mathsf{T}}S\mathcal{F} \right)_{\beta\alpha} \right|^2
\end{equation}

Expanding the matrix element representing the amplitude, $\text{Amp}_{\alpha\beta} = (\mathcal{F}STT^{\mathsf{T}}S\mathcal{F})_{\beta\alpha}$, and defining $J=TT^T$ we can write:
\begin{equation}
    \text{Amp}_{\alpha\beta} = \sum_{n,m}f_{\beta n}s_{nn}J_{nm}s_{mm}f_{m\alpha}
\end{equation}

This expression is a complex quadratic form in the phasor variables $s_{nn}\equiv s_n=e^{i\Phi_n}$. We can rewrite it compactly as:
\begin{equation}
    \text{Amp}_{\alpha\beta} = \sum_{n,m=1}^{N} A_{nm}^{(\alpha,\beta)} s_n s_m
    \label{eq:amp_2pds_SI_mid} 
\end{equation}
where the coupling matrix $A_{nm}^{(\alpha,\beta)} = f_{\beta n} J_{nm} f_{m\alpha}$ encapsulates the properties of the medium ($J=TT^T$) and the targeted modes $(\alpha, \beta)$.

The core optimization problem is thus to find the set of SLM phasors $s_n$ that maximizes the probability $P^{(2P-DS)}_{\alpha\beta} \propto |\text{Amp}_{\alpha\beta}|^2 = |\sum_{n,m} A_{nm}^{(\alpha,\beta)} s_n s_m|^2$.

Solving this optimization problem is computationally challenging due to the nature of the objective function and constraints. The task involves maximizing the magnitude of a complex quadratic form where the variables $s_n$ are coupled through the matrix $A^{(\alpha,\beta)}$ and are restricted to the unit circle in the complex plane ($|s_n|=1$). This structure shares features with computationally hard problems studied in other fields. For instance, it involves a quadratic objective function, central to Quadratic Programming, though the unit-modulus constraint is not identical to typical Quadratic Programming setups. Additionally, the expression $\sum A_{nm} s_n s_m$ is structurally similar to Hamiltonians used in spin glass models (e.g., the XY model if $A$ were real and symmetric). Finding ground states for such models, especially with disordered interactions $A_{nm}$, is often known to be NP-hard ~\cite{arora2005non, pardalos1991quadratic, lucas2014ising, pierangeli2019large, pierangeli2021scalable}. In fact, maximizing the magnitude squared $|\text{Amp}_{\alpha\beta}|^2$ in such a system has been mapped in ~\cite{courme2025non} to an Ising model involving multi-spin interactions. 

Although a formal complexity analysis for this specific problem, where $A^{(\alpha,\beta)}$ is derived from random matrices $T$ and subject to specific phasor constraints, is beyond the scope of this work, its structural resemblance to known hard problems suggests that finding the guaranteed global optimum is also computationally prohibitive for large system sizes $N$. The difficulty arises from the interdependence of all phase variables $s_n$ via the dense coupling matrix $A^{(\alpha,\beta)}$.

Despite these computational challenges, finding good approximate solutions numerically is feasible for moderate system sizes, as demonstrated by the gradient-based optimization results presented in the main text.

\section{Numerical simulations}\label{SI:sim-details}
In this section, we explain in detail our numerical simulations, used to find the expected enhancements for the 2P-DS configuration. We also use the simulation to verify that the results from Section \ref{SI:enhancement} hold similarly for random unitary matrices.

In the simulation, we compute the coincidence probabilities given by Eqs. (\ref{eq:Pab_1PS}), (\ref{eq:Pab_2PIS}), and (\ref{eq:Pab_2PDS}), and use a numerical optimizer to find the SLM phase mask that maximizes $P_{\alpha\beta}$. Each result shown in Fig. 3 of the main text was averaged from 200 disorder realizations. For each disorder realization we run the optimizer for the five configurations (1P-S, 2P-IS, 2P-IS(OPC), 2P-DS, and 2P-DS(OPC)), inserting the phase patterns at the appropriate plane and optimizing the appropriate (symmetric or non-symmetric) output mode.

The Gaussian IID matrices where generated using the `random.normal` function of the numpy module, normalized such that on average $\sum_{n=1}\left|t_{mn}\right|^{2}=1$. The random unitary matrices were generated using the `unitary\_group.rvs` function of the scipy.stats module. The dimensions of the matrices were set to $N=512$; see Section \ref{SI:N-dep} for further details.

In the following we provide a short walk-through of further simulation results, while the full details and code are provided in ~\cite{qwfs2024}. 

\subsection{Comparing optimizers for the 2P-DS configuration}
To optimize the phases in the 2P-DS configuration, we compared several optimizers. We found that simulated annealing, while well suited for finding global minima, did not scale well for systems with more than a few tens of modes, resulting in long run times and suboptimal performance. Next, we checked the scipy.minimize implementation of both the SLSQP and L-BFGS-B algorithms, and found that the L-BFGS-B method scaled better with the system size. 

Finally, we realized that using PyTorch’s autograd engine for simulations is advantageous, as it provides exact gradient computations and utilizes more information about the system. We found that both the ADAM and the lbfgs optimizers of pytorch performed well, and achieved better enhancements than the scipy implementations for large systems. In practice, all the numerical results in this paper were found using the pytorch lbfgs optimizer. The exact parameters used for the simulations, can be found in the code that we attach to this paper at Ref.~\cite{qwfs2024}.

\subsection{Validation of optimizer against analytical results}
\begin{figure}[ht!]
    \centering
    \includegraphics[width=\linewidth]{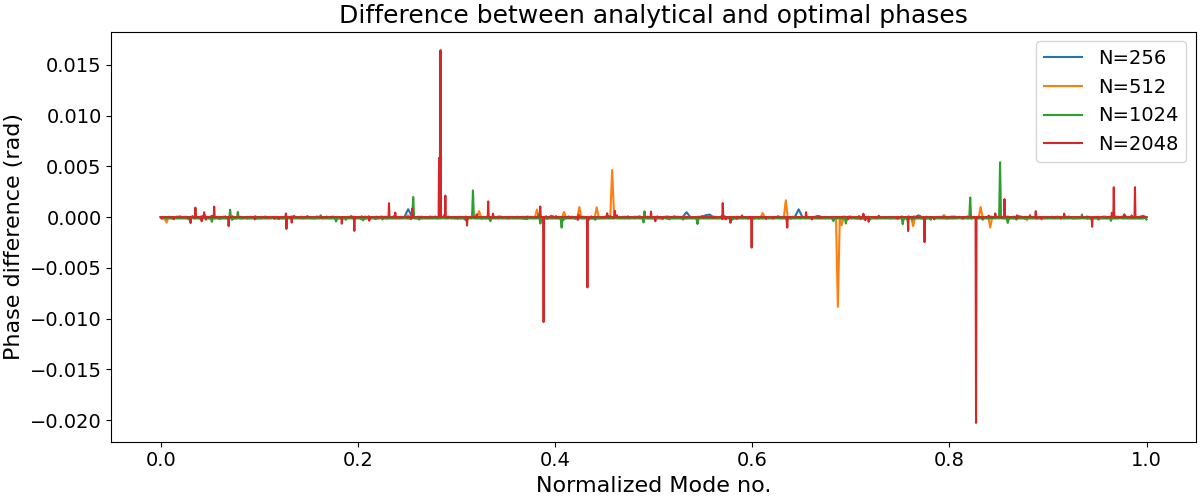}\caption{\label{fig:same_phases}Comparison of the phases found by the numerical optimizer and the analytical best phases. It can be seen that the difference between the numerical and analytical phases are very small, for various system sizes. The horizontal axis represents the different modes, where the mode numbers ($1$ to $N$) are rescaled to $[0,1]$.}
\end{figure}

To validate the performance of our gradient-based optimizer, we first tested it against configurations with known analytical solutions. We began with a standard classical focusing experiment through a scattering sample. The optimizer searches for the phases $\Phi_m$ that maximize the intensity at an output mode $\beta$, $I_{\beta}=|\sum_m t_{\beta m}e^{i\Phi_m}|^2$. We find that the phases found by the numerical optimizer closely match the analytical solution given by phase conjugation $\Phi_m=-\text{arg}(t_{\beta m})$, as illustrated in Fig.~\ref{fig:same_phases} for system sizes up to $N=2048$.

\begin{figure}[ht!]
    \centering
    \includegraphics[width=0.9\linewidth]{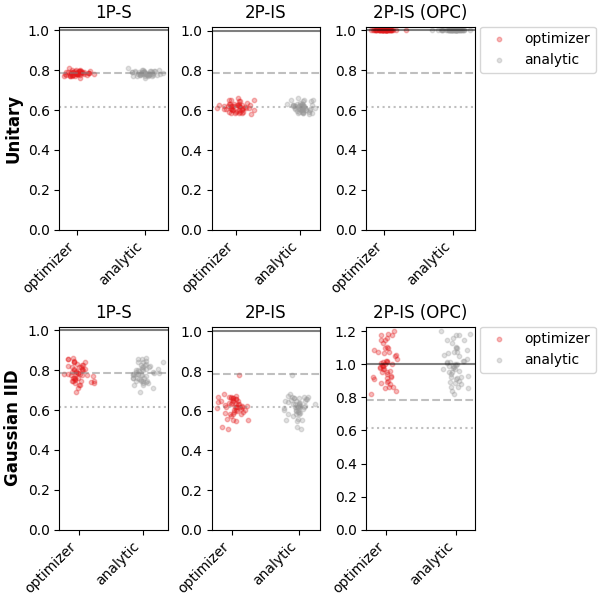} 
    \caption{\label{fig:sanity}Comparing the enhancement pre-factor $\eta/N$ obtained using phases optimized via the autograd lbfgs optimizer with the pre-factor obtained using analytically calculated optimal phases for the 1P-S, 2P-IS, and 2P-IS (OPC) configurations. The top row shows results for unitary matrices, and the bottom row for Gaussian IID cases. For each of 50 disorder realizations we show the achieved enhancement pre-factor, using either the analytical solution or the optimizer, where each mark represents a single run. It can be seen that the optimizer achieves enhancements very close to the analytical expectations. The horizontal lines represent the expected average enhancement factors for the 1P-S ($\frac{\pi}{4}$, dashed line), 2P-IS ($(\frac{\pi}{4})^2$, dotted line) and 2P-IS(OPC) (1, solid line) configurations.}
\end{figure}

We further tested the optimizer by comparing its numerical output with analytical predictions for the 1P-S and 2P-IS configurations. Figure~\ref{fig:sanity} shows the resulting enhancement pre-factors obtained for 50 disorder realizations (red markers). We compare these to the optimal enhancement achieved with analytically calculated optimal phases (conjugate phases) for the same disorder realizations (gray markers). Horizontal lines indicate the theoretical average values derived in Section \ref{SI:enhancement}. The enhancements obtained numerically by the optimizer consistently approach these predictions across disorder realizations, confirming the validity and robustness of the optimization results.

\subsection{2P-DS Gaussian IID example}
\begin{figure}[ht!]
    \centering
    \includegraphics[width=0.8\linewidth]{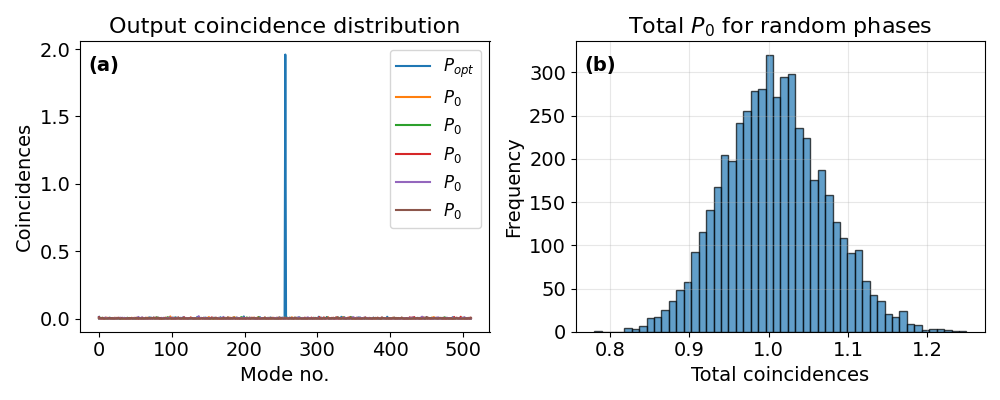} 
    \caption{\label{fig:SLM3_sample} (a) The output coincidence rate distribution for several realizations of random phases ($P_0$), and also for numerically found optimal phases ($P_{\text{opt}}$), where the coincidence rate at the optimized mode is shown to be larger than $1.9$ in this realization. (b) A histogram of the total output coincidences for 5000 realizations of random phases. The Gaussian IID matrices are generated such that on average, the total output coincidence rate before optimization is unity $N\cdot P_0=1$. }
\end{figure}

A notable feature of the 2P-DS configuration with Gaussian IID matrices is that the optimized coincidence rate can exceed the average total coincidence rate (summed over all output modes) observed before optimization. Figure \ref{fig:SLM3_sample}(a) shows the distribution of coincidence rate across output modes for several random SLM phase patterns and for an optimized phase pattern. The optimized pattern yields a strong peak at the target mode, with a normalized coincidence  rate that is nearly twice as high as the total rate obtained with a random SLM phase pattern. Figure \ref{fig:SLM3_sample}(b) shows a histogram of the total output coincidences (summed over all modes) for 5000 random phase patterns. The average total coincidence rate $N\cdot\langle P_0\rangle$ fluctuates with $N\cdot\langle P_0\rangle\approx1\pm0.1$. The optimized peak coincidence rate in the single target mode in Fig.~\ref{fig:SLM3_sample}(a) clearly exceeds this average total intensity, demonstrating an enhancement of the total transmission by a statistically significant amount. 

\subsection{2P-DS (OPC) configurtation: $0$ and $\pi$ phases}
\begin{figure}[ht!]
    \centering
    \includegraphics[width=\linewidth]{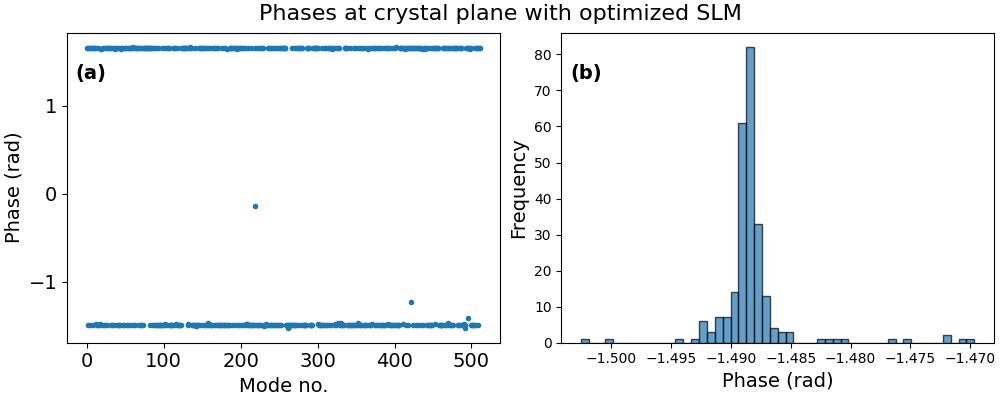} 
    \caption{\label{fig:zero_pi_phases}(a) The phases of the field in the crystal plane for an optimized 2P-DS (OPC) configuration in the unitary case. It can be seen that the phases are some combination of two phases, with a difference of $\pi$ between them. (b) A histogram of the phase values close to $-1.49$, showing quantitatively the clustering of the phases.}
\end{figure}

In the 2P-DS (OPC) configuration, for a unitary matrix, we explained in the main text that the empirically optimized phases are such that the phases in the crystal plane (mirror plane in the advanced wave picture) are some combination of two phases of the type $\Phi, \Phi + \pi$. In Fig. \ref{fig:zero_pi_phases} we show the phases of the field in the crystal plane after optimization. The phases clearly cluster around two values separated by $\pi$. A histogram of the phase values confirms this binary structure quantitatively (Fig. \ref{fig:zero_pi_phases}(b)).

\subsection{Enhancement dependence on size of system $N$ for the 2P-DS configuration}\label{SI:N-dep}
In standard wavefront shaping configurations, the enhancement pre-factor $\eta/N$ typically approaches a constant value for large $N$. For instance, $\eta_{1P-S}\approx\pi/4\cdot N$ for large $N$, with only $\mathcal{O}(1/N)$ corrections. 

To understand the $N$ dependence in the Gaussian IID case, we first notice that when optimizing the coincidences in a specific output mode, two things are optimized simultaneously: a) The total amount of transmitted energy, which is determined by the overlap of the incident wave with the transmission eigenchannels with large transmission eigenvalues, and b) The energy in the specific optimized mode. Pictorially, the optimal phases find a way to harvest energy from the fluctuations of the TM. This is reminiscent of Ref. ~~\cite{vellekoop2008universal}, where it has been shown in a classical configuration that the optimal phases that focus light through a real scattering medium also excite the open channels and enhance the total transmitted power. In our case, the maximal energy in the channel is bounded from above by $\sigma_{1}^2$ for $\sigma_{1}$ the largest singular value of $TT^T$. 

\begin{figure}[ht!]
    \centering
    \includegraphics[width=\linewidth]{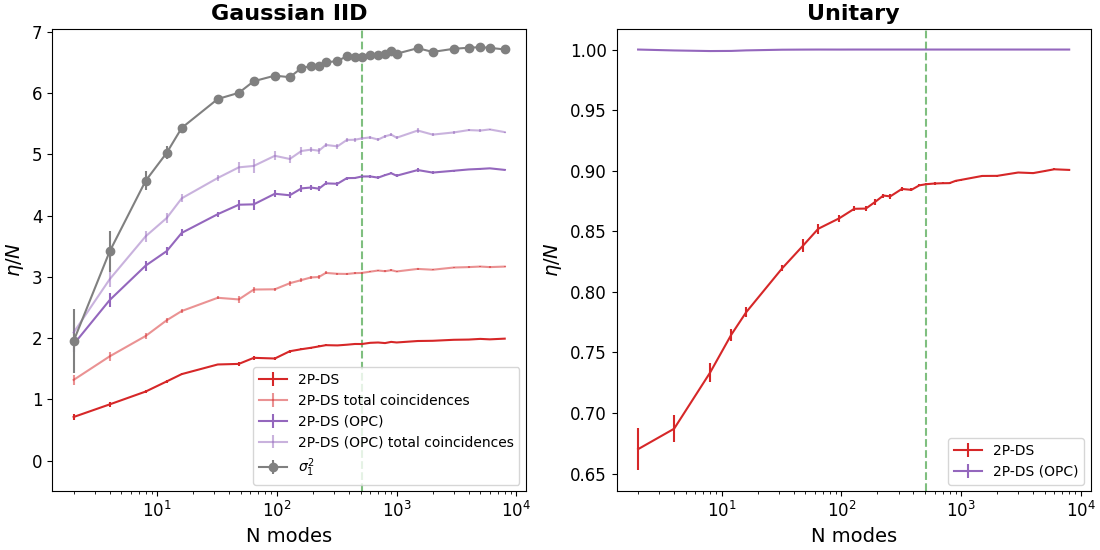} 
    \caption{\label{fig:N_dependence} Dependence of the enhancement pre-factor ($\eta/N$) on system size $N$ (log scale) for the 2P-DS configuration (Unitary and Gaussian IID, symmetric and non-symmetric cases). For the Gaussian IID case we also show the squared largest singular values $\sigma_{1}^2$ which bound from above the total optimized coincidence counts, which is also shown. The slopes of $\sigma_{1}^2$, the total coincidence counts, and the optimized enhancement seem to change similarly, hinting at a connection between them. The dashed green line indicates $N=512$ which is the dimension we used in the rest of this work. Error bars represent the standard deviation over $k$ realizations, where $k$ varies between $300$ for small $N$ and $5$ for large $N$.}
\end{figure}

To numerically investigate the dependence of the enhancement pre-factor on the system size $N$, we performed simulations of the 2P-DS configuration across a range of $N$ values. Fig. \ref{fig:N_dependence} plots the optimized enhancement pre-factor $\eta/N$ for different values of $N$ for both Gaussian IID and Unitary matrices, considering both symmetric (OPC) and non-symmetric detection. The simulations show that for the Gaussian IID case, in both detection schemes, $\eta/N$ increases significantly at small $N$ and then exhibits a tendency towards saturation at $N$ values of a few hundred. Fig. \ref{fig:N_dependence} also shows $\sigma_{1}^{2}$, and the total optimized coincidence counts, which both saturate at similar values of $N$. However, there does seem to be a continued non-negligible dependence on $N$ (larger than $\mathcal{O}(1/N)$) even for a few thousand modes. The results presented in the main text were computed for $N=512$ modes, where $\eta/N$ has began to saturate, and the optimization converges within reasonable time. 

Interestingly, a similar trend of increasing $\eta/N$ with slow saturation is observed also for unitary matrices in the non-symmetric case (Fig. \ref{fig:N_dependence}), even though  the total transmission is inherently fixed to unity and all singular values of $TT^T$ are $1$. This indicates that the $N$-dependence in the unitary case cannot be attributed to varying transmission efficiency. In the symmetric unitary case the enhancement pre-factor is constant at unity. 

\section{2P-IS configuration with the SLM not in the crystal plane}\label{SI:SLM5}
In the main text, the analysis of the two-photon illumination shaping (2P-IS) configuration assumes the SLM is located precisely in the image plane of the nonlinear crystal. This ensures both photons acquire the exact same phase modulation $S$. In the advanced wave picture, this corresponds to a classical wave propagating backwards through the scattering medium, hitting the SLM, hitting it immediately again, and then propagating forwards through the medium again, as described in Eq.~ (\ref{eq:Pab_2PIS}) by $\mathcal{F}TSST^T\mathcal{F}$. 

If the SLM is displaced from the crystal's image plane, the advanced wave picture becomes slightly more complex. The classical wave, after its first interaction with the SLM, undergoes further propagation (corresponding to the path from the SLM to the crystal plane and back) before impinging on the SLM a second time. Let $P$ represent the operator for this round-trip propagation between the two SLM interactions. The propagation operator would then be $\mathcal{F}TSPST^T\mathcal{F}$. Since the operator $P$ includes diffraction, the phase applied by a given SLM pixel during the first interaction influences the field impinging on other SLM pixels during the second interaction. This creates interdependencies between the SLM pixels, resulting in a self-consistent optimization problem similar to the 2P-DS configuration, which is generally difficult to solve.

To gain insight into the effects of such a displacement, we numerically investigate a scenario representing strong mode mixing between the two SLM interactions. Specifically, we model the intermediate propagation operator by the Fourier operator $P=\mathcal{F}$. We optimize the SLM phase mask numerically similar to the 2P-DS configuration (see section \ref{SI:sim-details}), and analyze the resulting enhancement in both the symmetric ($\alpha=\beta$), and non-symmetric ($\alpha\neq\beta$) cases. 

In the symmetric case, we once again achieve perfect enhancement $\eta=N$, similar to the case with no free-space propagation ($P=I$). This could be understood similar to the 2P-DS (OPC) configuration, where the SLM finds phases such that at the crystal plane the phases will be some combination of $0$ and $\pi$. 
 
In the non-symmetric case the opitmization yields an enhancement of $\eta\approx 0.852\cdot N$, which is better than both the 1P-S and 2P-IS results. Similar to the 2P-DS case, the fact that the SLM pixels become interdependent makes the optimization problem harder, but allows for better enhancement compared to the case with no free-space propagation. 

However, unlike the 2P-DS case, where the two interaction with the SLM occur once before and once after the scattering ($\mathcal{F}STT^TS\mathcal{F}$), here both interactions with the SLM occur within the scattering process ($\mathcal{F}TS\mathcal{F}ST^T\mathcal{F}$). Consequently, the SLM cannot optimize the total transmitted power in the Gaussian IID case. Indeed, the reported enhancements hold similarly for both unitary and Gaussian IID transmission matrices. 

In future work, it would be of interest to perform a full-wave simulation of these configurations, and systematically explore the dependence of the enhancement on the free-space propagation distance, represented by $P$. 

\section{Incomplete control}\label{SI:incomplete}
In this section, we compute the optimal enhancements for the 1P-S and 2P-IS configurations, assuming incomplete control. To this end, we compute the optimal probability $P_{\text{opt}}$ assuming a system with $N$ modes yet only $M<N$ independent \textit{macro-pixels} on the SLM, such that every macro-pixel is illuminated by $m=\frac{N}{M}$ modes. For $P_0$, we use Eq.~(\ref{eq:P0_T}) and Eq.~(\ref{eq:P0_TT}) derived in section \ref{SI:enhancement}, since the probability before optimization is independent of the degree of control.

\subsection{1P-S configuration}
In the 1P-S configuration, we take the right hand side of Eq. (\ref{eq:sum_1PS}), but now we must introduce a double summation, since there are less SLM pixels than modes:

\begin{equation}
    P_{\alpha\beta}=\frac{2}{N}\left|\sum_{n}h_{\beta n}^{(2)}h_{\alpha n}^{(1)}\right|^{2}=\frac{1}{2N}\left|\sum_{i=0}^{M-1}e^{i\Phi_{i}}\sum_{j=1}^{m}f_{\alpha\left(im+j\right)}t'_{\beta\left(im+j\right)}\right|^{2}.
\end{equation}

When the SLM is optimized, the phases $\Phi_{i}$ are chosen such that $\sum_{j=1}^{m}f_{\alpha\left(im+j\right)}t'_{\beta\left(im+j\right)}=\left|\sum_{j=1}^{m}f_{\alpha\left(im+j\right)}t'_{\beta\left(im+j\right)}\right|$ which for convenience we will denote $\left|x_{i}\right|$. Plugging this in we get:

\begin{equation} \label{eq:xj}
    P_{\text{opt}} =\frac{1}{2N}\left\langle \left|\sum_{i=0}^{M-1}\left|x_{i}\right|\right|^{2}\right\rangle = \frac{M}{2N}\left\langle \left|x_{i}\right|^{2}\right\rangle +\frac{M\left(M-1\right)}{2N}\left\langle \left|x_{i}\right|\right\rangle ^{2}
\end{equation}

In this case, $x_i=\sum_{j=1}^{m}f_{\alpha\left(im+j\right)}t'_{\beta\left(im+j\right)}$ is a sum of $m$ circular complex Gaussian variables, so it is such a circular Gaussian variable, with a variance
$\left\langle \left|x_{j}\right|^{2}\right\rangle =\frac{m}{N}\left\langle \left|t'_{ij}\right|^{2}\right\rangle$, with the factor of $1/N$ resulting from the fact that$|f_{ij}|=1/\sqrt{N}$. We thus get $\left\langle \left|x_{j}\right|\right\rangle^{2}=\frac{\pi}{4}\left\langle \left|x_{j}\right|^{2}\right\rangle=\frac{\pi}{4} \frac{m}{N}\ \left\langle \left|t'_{ij}\right|^{2}\right\rangle$. Dividing Eq.~(\ref{eq:xj}) by $P_0=\frac{1}{2N}\langle\!\langle|\tilde{\tilde{t}}_{\beta\alpha}|^2\rangle\!\rangle=\frac{1}{2N}\langle\!\langle|t'_{\beta\alpha}|^2\rangle\!\rangle$ [Eq.~(\ref{eq:P0_TT})] and using the fact that $mM=N$ we get:

\begin{equation}
    \eta_{1P-S}^{(i)} = 1+\left(M-1\right)\frac{\pi}{4}.
\end{equation}

We obtain that in the 1-PS configuration, the enhancement scales linearly with the amount of macro-pixels, with a $\pi/4$ pre-factor, as expected classically. 

\subsection{2P-IS configuration}
In the 2P-IS configuration, a correct choice of phases on the SLM will result again with Eq.~(\ref{eq:xj}), where now $x_i=\sum_{j=1}^m \tilde{t}_{\beta(im+j)}\tilde{t}_{\alpha(im+j)}$. However, since now $x_i$ is a sum of products of circular Gaussian random variables, $\left\langle \left|x_{i}\right|\right\rangle$ does not have a closed analytical expression. 

For $m\gg1$, corresponding to a large number of modes per macro-pixel and thus a low degree of control, we may use the central limit theorem. The variance of $x_i$ is thus given by $\langle|x_i|^2\rangle=m\langle|\tilde{t}_{\beta n}\tilde{t}_{\alpha n}|^2\rangle=m\langle|\tilde{t}_{\alpha n}|^2\rangle^2$, where in the last step we assumed $\alpha\neq \beta$. By the central limit theorem, $x_i$ is also a complex circular Gaussian variable, thus $\langle|x_i|\rangle=\frac{\pi}{4}\langle|x_i|^2\rangle=\frac{\pi}{4}m\langle|\tilde{t}_{\alpha n}|^2\rangle^2$. Plugging these values to Eq.~(\ref{eq:xj}) and dividing by $P_0=\langle\!\langle|\tilde{t}_{\alpha n}|^2\rangle\!\rangle^2$ [Eq.~(\ref{eq:P0_T})], we get:

\begin{equation}
\label{eq:eta_2p_partial}
    \eta_{2P-IS}^{(i)} = 1+\left(M-1\right)\frac{\pi}{4}.
\end{equation}

This agrees with the numerical results shown in Fig. \ref{fig:small_control}. An intuitive explanation of this result can be obtained using the advanced wave picture, where, in principle, for the 2P-IS configuration, the SLM is illuminated by a speckle field, leading to the $\left( \frac{\pi}{4} \right)^2$ prefactor in the full control limit [Eq.~(\ref{eq:eta_2ps})]. However, in the low degree of control limit $(m\gg  1)$, the total intensity illuminating each macro-pixel becomes approximately uniform. As a result, the SLM effectively sees a homogeneous illumination, yielding a prefactor of $\frac{\pi}{4}$ in Eq.~(\ref{eq:eta_2p_partial}), as in the 1P-S configuration [Eq.~(\ref{eq:eta_1ps})].

\subsubsection*{Digital Optical Phase Conjugation (OPC)}

When the two detectors are in the same transverse position ($\alpha=\beta$), we find $\langle|x_i|^2\rangle=m\langle|\tilde{t}_{\alpha n}|^4\rangle=2m\langle|\tilde{t}_{\alpha n}|^2\rangle$, where in the last step we used the Gaussian statistics of $\tilde{t}_{\alpha n}$. We then obtain:

\begin{equation}
    \eta_{2P-IS~(OPC)}^{(i)} =2+\left(M-1\right)\frac{\pi}{2}.
\end{equation}

For $M=1$, which corresponds to applying only a global phase and thus offering no effective control over the system, we get on average an enhancement of $2$, which is the known coherent back-scattering (CBS) two-fold enhancement. The resulting slope of $\pi/2$ which is two-fold larger than the classical $\pi/4$ may be seen as an extension of the CBS two-fold enhancement from the regime of no control to the regime of weak control. 

\subsection{2P-DS configuration}
\begin{figure}[ht!]
    \centering
    \includegraphics[width=\linewidth]{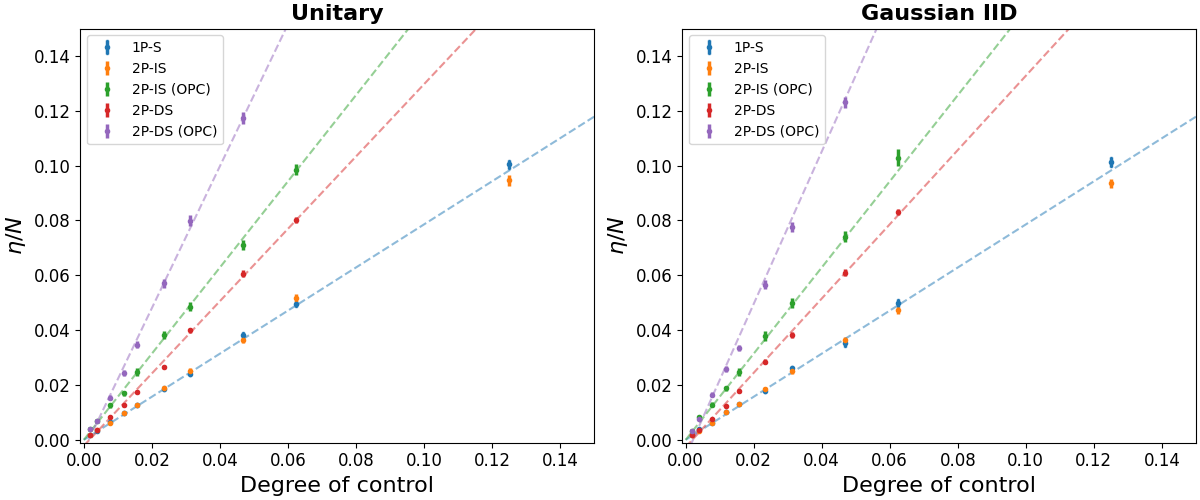} 
    \caption{\label{fig:small_control} The dependence of the enhancement pre-factor ($\eta/N$) obtained from numerical simulations, on the degree of control in the weak control regime. In this regime, the 1P-S and 2P-IS configurations both have a slope of $\pi/4$, and the 2P-IS (OPC) configuration has a slope of $\pi/2$, as shown in analytically drawn dashed lines. For the 2P-DS configurations, the dashed lines represent linear fits.}
\end{figure}

While the enhancement $\eta$ for the 2P-DS configuration lacks a simple analytical form, we numerically analyzed its scaling in the regime of weak control. As shown in Fig.~\ref{fig:small_control}, in this limit the enhancement pre-factor $\eta/N$ scales linearly with the degree of control. 

Linear fits to the simulation data yield the slopes at low degree of control. For unitary matrices, the slope is $\approx 1.32$ for non-symmetric detection ($\alpha \neq \beta$) and $\approx 2.58$ for symmetric detection ($\alpha = \beta$). For Gaussian IID matrices, the corresponding slopes are $\approx 1.35$ and $\approx 2.78$. 

Notably, for both matrix types, the initial slope for the symmetric 2P-DS(OPC) case is approximately twice that of the non-symmetric case, mirroring the factor-of-two difference between the analytical initial slopes of 2P-IS(OPC) and standard 2P-IS ($\pi/2$ vs $\pi/4$).

\begin{figure}[ht!]
    \centering
    \includegraphics[width=\linewidth]{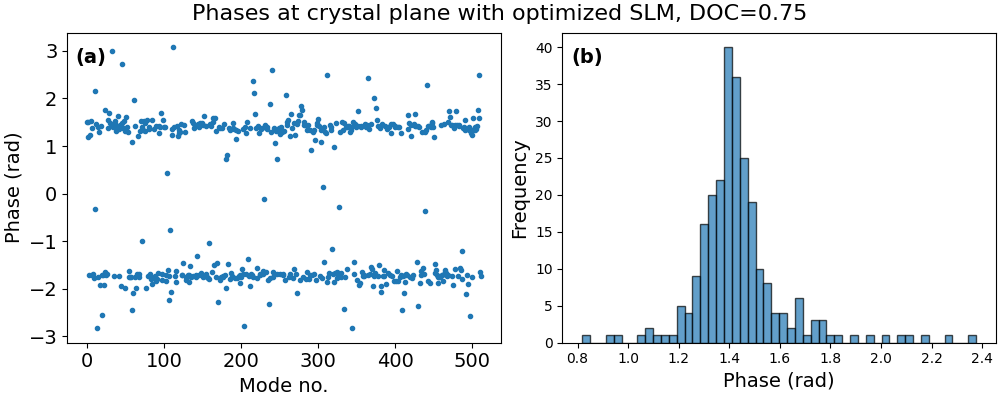} 
    \caption{\label{fig:zero_pi_phases_incomplete_control}(a) The phases of the field in the advanced wave picture of the 2P-DS (OPC) configuration, obtained in the crystal plane. In this example, we assume the scattering sample is described by a unitary matrix, and that the degree of control (DOC) is $0.75$. The passes exhibit a bimodal distribution, clustering around two phases separated by $\pi$. (b) A histogram of the phase values near $1.4$, quantitatively demonstrating phase clustering.}
\end{figure}

In the main text, we explained that for scattering samples described by unitary matrices, since there are $2^N$ OPC-like solutions, even when control is incomplete some of the solutions are approximately reachable. To demonstrate this, in Fig. \ref{fig:zero_pi_phases_incomplete_control} we show the phases of the field in the advanced wave picture, obtained at the crystal plane after optimization, for a degree of control of $0.75$. While less sharply defined than with full control, the phases still exhibit a strong tendency towards a binary distribution, consistent with achieving near-optimal enhancement even under incomplete control. 

\subsection{Thin diffuser limit}\label{SI:thin}
The main text focuses on thick scattering media described either by random unitary matrices, or by Gaussian IID matrices where the elements of the matrix are independently and identically distributed. At the other extreme, a thin scattering sample can be modeled as a single phase mask, representing a thin diffuser, described mathematically by a diagonal matrix $t_{ij}=\delta_{ij}e^{i\phi_i}$ where $\delta_{ij}$ is the Kronecker delta. Since the SLM is also described by a diagonal matrix, and diagonal matrices commute, all proposed configurations yield the exact same results. 

\begin{figure}[ht!]
    \centering
    \includegraphics[width=0.85\linewidth]{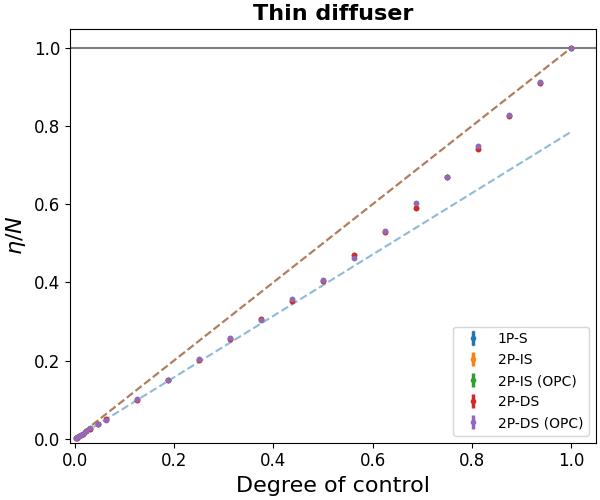} \caption{\label{fig:thin_diffuser} Dependence of the enhancement pre-factor $\eta/N$ on the degree of control for a thin diffuser. All configurations yield identical results, transitioning from a $\pi/4$ slope at low control to perfect enhancement ($\eta/N=1$) at full control. Dashed linear lines with slopes of of $\pi/4$ and $1$ are intended to guide the eye for this transition.}
\end{figure}

With full control, as depicted in Fig. \ref{fig:thin_diffuser}, the SLM perfectly compensates the scattering by applying the conjugate phases $-\phi_i$. In the regime of limited control, we obtain the classical result with a slope of $\pi/4$. In future work, it would be of interest to study the different configurations in the intermediate regime of a moderately thick diffuser.

\bibliography{SI}